Banner appropriate to article type will appear here in typeset article

# General model for shear stress and surface pressure distributions for flow over spheres

**Bryce D. Daniels[1] † and Thomas E. Schwartzentruber[1]**

[1]Department of Aerospace Engineering and Mechanics, University of Minnesota, Minneapolis, MN 55455, USA

A general physics-based model for the distribution of shear stress and surface pressure is developed for laminar flow over spheres. Although integral coefficients have been studied extensively for spheres, they do not capture the local surface distributions that govern phenomena such as heat transfer, phase change, and mass loss in multiphase and particle-laden flows. The model depends on the flow regime of the sphere ($Re_\infty$, $M_\infty$, $Kn_\infty$), local gas properties ($\gamma$, $\omega$, $Pr$), and surface conditions ($T_w/T_0$, $\sigma_t$, $\sigma_n$). The formulation incorporates boundary-layer scaling, flow separation, shock-wave physics, velocity-slip effects, and high-speed rarefaction, while recovering the known asymptotic behavior in the creeping-flow, hypersonic, and free-molecular limits. The proposed model is validated in the continuum and rarefied regimes using CFD and DSMC simulations from literature. In the continuum regime, it is compared with CFD simulations of flow over a sphere in air under subsonic, supersonic, and hypersonic conditions with varying surface temperatures. The model captures the CFD distributions very well over a wide range of $Re_\infty$, $M_\infty$, and $T_w/T_0$. In the rarefied regime, it is compared with DSMC simulations of flow over a sphere in a monatomic gas under subsonic and supersonic conditions with fully diffuse and partial surface accommodation. It shows good agreement with the simulated shear stress and surface pressure distributions over a wide range of $M_\infty$ and $Kn_\infty$, while accurately capturing the effects of surface accommodation around spherical bodies.

## 1. Introduction

Viscous flow over a sphere has been investigated extensively through experimental, theoretical, and numerical studies. Although the sphere is the simplest three-dimensional shape, the flow field involves complex physical phenomena such as flow separation, recirculation, vortex shedding, and boundary-layer transition, which all depend on the flow conditions (Tandea 1956; Johnson & Patel 1999; Rodriguez *et al.* 2011; Achenbach 1972). In the continuum regime, the flow behavior is commonly categorized by the freestream Reynolds number ($Re_\infty$). For example, flow separation and wake unsteadiness occur for $Re_\infty$ above approximately 20 and 270, respectively. Additionally, as $Re_\infty$ approaches 2.5 x $10^5$, the boundary layer begins to transition from laminar to turbulent prior to separation, which corresponds to the critical transition condition. As the Mach number ($M_\infty$) and Knudsen number ($Kn_\infty$) increase, compressibility and rarefaction effects can significantly alter the flow structure (Nagata *et al.* 2016, 2020*a*; Holman & Boyd 2008).

† Email address for correspondence: dani0588@umn.edu

The structure of the flow directly affects the surface quantities, including the shear stress ($\tau_w$) and surface pressure ($p_w$), which dictate the aerodynamic forces acting on the body. For most engineering applications, integral quantities such as the drag coefficient are used to model the motion of spheres in fluid flows; however, they do not capture the key physical effects occurring along the surface. The distributions of $\tau_w$ and $p_w$ are particularly useful for multiphase and particle-laden flow applications, where heat transfer, phase change, and particle deformation depend on the local surface conditions. For example, in high-speed flows, such as planetary entry, vehicles can encounter clouds of dust particles, ice crystals, and water droplets that absorb energy from the high-temperature post-shock flow and may undergo sublimation, melting, or breakup prior to reaching the surface (Habeck 2023; Dworzanczyk *et al.* 2025). Additionally, since the particles may encounter a wide range of flow conditions, including both the continuum and rarefied regimes, it is convenient to express $\tau_w$ and $p_w$ in terms of the dimensionless skin friction ($C_f$) and surface pressure ($C_p$) coefficients, respectively. These coefficients are defined as

$$C_f = \frac{\tau_w}{1/2\rho_\infty U_\infty^2} \tag{1.1}$$

$$C_p = \frac{p_w - p_\infty}{1/2\rho_\infty U_\infty^2} \tag{1.2}$$

where $U_\infty$, $\rho_\infty$, and $p_\infty$ correspond to the freestream velocity, density, and pressure, respectively. Theoretical models for $C_f$ and $C_p$ have been developed for specific flow conditions. In the continuum regime ($Kn_\infty < 0.01$), under incompressible creeping-flow conditions ($M_\infty < 0.3$, $Re_\infty \ll 1$), analytical solutions for $C_f$ and $C_p$ can be obtained from the Navier-Stokes equations (Stokes 1851). Similarly, under high-speed inviscid conditions ($M_\infty \gg 1$, $Re_\infty \gg 1$), a closed-form expression for the surface pressure coefficient can be derived by considering the change in the surface-normal momentum flux (Newton 1687). For $Kn_\infty > 10$, gas molecules travel large distances prior to colliding and the flow around the sphere is considered to be free-molecular. In the free-molecular limit, analytical expressions for $C_f$ and $C_p$ can be derived by considering the momentum and energy exchange of incident and reemitted gas molecules with the surface (Schaaf & Chambre 1958). Physics-based models are needed to accurately predict the distributions of $C_f$ and $C_p$ across the continuum and rarefied regimes, while also recovering the limiting behavior. To achieve this, the primary physical effects of each regime must be incorporated into the model formulation.

As $Re_\infty$ increases and the velocity gradient at the surface decreases to zero, the flow separates from the sphere and the distribution of $C_f$ becomes confined between the stagnation and separation points. Additionally, in the afterbody region of the sphere, the distribution of $C_p$ is influenced by the length of the recirculation region. At high $Re_\infty$, a boundary layer develops along the surface, and the skin friction distribution depends on the acceleration of the external flow and variation of the displacement and momentum thicknesses around the sphere. These effects are captured through scaling laws derived from the momentum-integral equation. The surface pressure distribution is primarily controlled by the inviscid conditions at the edge of the boundary-layer, flow separation, and the stability of the wake. For compressible flow ($M_\infty \geqslant 0.3$), density variations become non-negligible, and the distributions of $C_f$ and $C_p$ are modified by compressibility and surface-temperature effects. When the flow becomes supersonic ($M_\infty \geqslant 1$), the detached bow shock rapidly compresses the gas in front of the sphere, resulting in a large jump in the flow properties. Additionally, the shock standoff distance effects the rate of acceleration of the flow on the forebody of the sphere. In the rarefied regime ($Kn_\infty > 0.01$), fewer collisions occur between gas molecules, and the assumptions of continuum flow begin to break down. Within the slip regime ($0.01 \leqslant Kn_\infty < 0.1$), the Navier-Stokes equations can be derived using the first-order Chapman-Enskog solution of the Boltzmann equation (Chapman & Cowling 1970). The surface boundary conditions that are consistent with this derivation include the velocity slip and temperature jump conditions (Maxwell 1879; Kennard 1938). At higher $Kn_\infty$, rarefaction scaling parameters for the surface quantities can be formulated by considering higher-order corrections to the Chapman-Enskog solution (Singh & Schwartzentruber 2016; 2017).

The current state of the art for determining the distributions of $C_f$ and $C_p$ relies on the use of computational methods to fully resolve the flow field around the sphere. This requires computational fluid dynamics (CFD) and direct simulation Monte Carlo (DSMC) simulations to numerically solve the Navier-Stokes and Boltzmann equations, respectively. However, both methods can be computationally expensive and are not practical for simulating a large number of spherical particles that experience a wide range of flow conditions along a trajectory. Therefore, the objective of this work is to develop general analytical models for the local shear stress and surface pressure around spheres that span a wide range of $Re_\infty$, $M_\infty$, and $Kn_\infty$, while also capturing the primary effects of each flow regime.

The final set of model equations are listed in the Appendix. Sections § 2–3 of this article are arranged to present the derivation of and physical rationale for the model functional form in each flow regime. The model free parameters, all in the continuum regime, are then prescribed in § 4 using CFD data from literature. Finally, the model is compared to shear stress and surface pressure distributions from fully resolved CFD and DSMC simulations in § 5–6 and conclusions are summarized in § 7.

## 2. Continuum regime

### 2.1. *Incompressible flow ($M_\infty < 0.3$)*

At low Reynolds numbers ($Re_\infty \lll 1$), the inertial terms in the momentum balance are neglected and the steady state Navier-Stokes equations can be reduced to a set of linearized equations, known as the Stokes equations (Stokes 1851). The simplified set of equations can be analytically solved for uniform flow over a sphere, resulting in the following expression for the skin friction coefficient

$$C_f = \frac{6}{Re_\infty} \sin\theta \tag{2.1}$$

where $\theta$ is the polar angle measured from the stagnation point. When $Re_\infty$ becomes sufficiently large ($\gg 1$) and the inertial forces in the flow dominate, viscous effects are confined within a thin boundary-layer surrounding the sphere. Within this region, the Navier-Stokes equations can be simplified using boundary-layer approximations. By integrating these equations from the surface of the sphere to the boundary-layer edge, the momentum-integral equation is obtained, which relates the surface shear stress to variations in the streamwise momentum flux and pressure gradient. For a sphere, the momentum-integral equation can be expressed in terms of $C_f$ with

$$C_f = \frac{\rho_e U_e^2}{\rho_\infty U_\infty^2} \frac{2}{R} \frac{d\delta_2}{d\theta} + \frac{\rho_e U_e}{\rho_\infty U_\infty^2} \frac{2}{R} \frac{dU_e}{d\theta} (2\delta_2 + \delta_1) + \frac{\rho_e U_e^2}{\rho_\infty U_\infty^2} \frac{2}{R} \frac{\delta_2}{r} \frac{dr}{d\theta} \tag{2.2}$$

where $U_e$ is the boundary layer edge velocity, $\rho_e$ is the boundary layer edge density, $R$ is the radius of the sphere, $r = R\sin\theta$, and $\delta_1$ and $\delta_2$ correspond to the displacement and momentum thickness, respectively. Since a simple closed-form expression for $C_f$ does not exist in this regime, historical approaches have involved numerically solving (2.2) up to the point of boundary layer separation. Instead, it is convenient to approximate the scaling of $C_f$ in terms of the global boundary layer variables. By assuming the flow at edge of the boundary layer is inviscid, the potential flow solution for $U_e$ can be used ($U_e = 3/2U_\infty \sin\theta$). Additionally, $\delta_1$ and $\delta_2$ can be approximated using the boundary layer thickness ($\delta$). Therefore, for incompressible flow over a sphere, the scaling of $C_f$ in (2.2) can be simplified as

$$C_f \sim \frac{d(\delta/R)}{d\theta}\sin^2\theta + \frac{\delta}{R}\sin(2\theta) \tag{2.3}$$

where $\delta \sim \sqrt{\mu s/\rho U}$ and $s = R\theta$. After substituting the potential flow expression for $U_e$ into $\delta$ and simplifying, the scaling of the nondimensional boundary layer thickness ($\delta/R$) becomes

$$\frac{\delta}{R} \sim Re_\infty^{-1/2}\left(\frac{\theta}{\sin\theta}\right)^{1/2} \tag{2.4}$$

On the forebody region of the sphere, the angular dependence in (2.4) remains close to unity since $\sin\theta \approx \theta$ for small to moderate $\theta$. Thus, the rate of change of $\delta/R$ is relatively small, and $C_f$ can be approximated by the second term in (2.3), $C_f \approx C_0 Re_\infty^{-1/2}\sin(2\theta)$. Based on the boundary layer calculations of Yuge (1956), prior to the critical transition condition, $C_0 = 3.8$ for incompressible flow over a sphere. By combining this approximation with (2.1), a general expression for $C_f$ on the forward portion of the sphere can be written as

$$C_f = C_{f,\max}\sin\left(\frac{\pi}{2}\frac{\theta}{\theta_{\max}}\right) \tag{2.5}$$

where $C_{f,\max}$ and $\theta_{\max}$ denote the maximum skin friction coefficient and the corresponding angular position, respectively. These terms are both functions of $Re_\infty$ that maintain the correct limiting scaling behavior.

As the flow approaches the shoulder of the sphere, the surface pressure reaches a minimum value (hence $dU_e/d\theta = 0$), which is captured by $\sin(2\theta) \approx 0$ in the second term of (2.3). Additionally, upon differentiating the angular terms in (2.4), the result remains close to unity near the shoulder, so only the dependence of $Re_\infty^{-1/2}$ is retained. Therefore, $C_f \sim Re^{-1/2}\sin^2\theta$ around the shoulder of the sphere. Since the flow can separate from the surface in this region, a more general expression is required that captures this effect, which occurs for $Re_\infty > 20$. At the point of boundary layer separation ($\theta_s$), the velocity gradient normal to the surface (and therefore $\tau_w$) is zero. Assuming the wall shear stress remains very small downstream of $\theta_s$, a general piecewise expression for $C_f$ is

$$\begin{aligned}
C_f &= C_{f,\max}\sin\left(\frac{\pi}{2}\frac{\theta}{\theta_{\max}}\right) && 0^\circ \leqslant \theta \leqslant \theta_{\max} \\
C_f &= C_{f,\max}\sin^2\left(\frac{\pi}{2}\frac{\theta-\theta_s}{\theta_{\max}-\theta_s}\right) && \theta_{\max} < \theta \leqslant \theta_s \\
C_f &= 0 && \theta_s < \theta \leqslant 180^\circ
\end{aligned} \tag{2.6}$$

where $\theta_s$ is a known function of $Re_\infty$ for incompressible flow (refer to (A 30)).

Following a similar approach to the skin friction coefficient, a general expression for $C_p$ can be determined by first considering fully attached flow. In the limits of low and high $Re_\infty$, analytical expressions for $C_p$ are obtained from Stokes and potential flow, respectively. These expressions can be written as

$$C_p = -\frac{6}{Re_\infty} + \frac{12}{Re_\infty}\cos^2\left(\frac{\theta}{2}\right) \tag{2.7a}$$

$$C_p = -\frac{5}{4} + \frac{9}{4}\cos^2\theta \tag{2.7b}$$

In both expressions, the leading coefficient represents the minimum value ($C_{p,\mathrm{min}}$), and the second coefficient represents the total peak-to-peak amplitude, which is the total amplitude between the stagnation point ($C_{p,0}$) and $C_{p,\mathrm{min}}$. Therefore, a general expression for $C_p$ in fully attached flow is given by

$$C_p = C_{p,\mathrm{min}} + \left(C_{p,0} - C_{p,\mathrm{min}}\right)\cos^2\left(\frac{\pi}{2}\frac{\theta}{\theta_{\mathrm{min}}}\right) \tag{2.8}$$

where $\theta_{\mathrm{min}}$ is the angle corresponding to $C_{p,\mathrm{min}}$. The value of $C_{p,0}$ is strongly influenced by viscous effects at low $Re_\infty$. Homann (1952) developed a viscosity correction factor for $C_{p,0}$ in incompressible flow by considering the velocity gradient and displacement thickness at the point of incidence which can be generalized with

$$C_{p,0} = C_{p,0,M} + \frac{12}{Re_\infty + 0.644\sqrt{Re_\infty}} \tag{2.9}$$

where $C_{p,0,M}$ is the inviscid contribution due to the compressibility of the flow. Expressions for $C_{p,0,M}$ will be derived in the following subsections for subsonic and supersonic flow.

Before the flow separates from the surface of the sphere, it experiences an adverse pressure gradient downstream of $\theta_{\mathrm{min}}$. After separation occurs, a recirculation region forms in the wake of the sphere. Within this region, the surface pressure distribution over the base of the sphere remains nearly uniform, which corresponds to the base pressure coefficient ($C_{p,\mathrm{b}}$). Since the surface pressure continues to increase downstream of $\theta_s$ before leveling out near the base ($\theta_{\mathrm{b}}$), it is convenient to approximate the scaling of $C_{p,\mathrm{b}}$ using the total pressure increase from $C_{p,\mathrm{min}}$. Therefore, an expression for $C_{p,\mathrm{b}}$ can be approximated as

$$C_{p,\mathrm{b}} = C_{p,\mathrm{min}} + \Delta C_{p,r} \tag{2.10}$$

where $\Delta C_{p,r}$ is the total pressure recovery downstream of $\theta_{\mathrm{min}}$. The value of $\Delta C_{p,r}$ primarily depends on the structure of the separated flow, including the location of the separation point, recirculation length, and wake stability, which all vary with $Re_\infty$. Additionally, to maintain the correct scaling in Stokes flow, $\Delta C_{p,r}$ should reduce to zero for $Re_\infty \ll 1$. A general piecewise expression for $C_p$ can now be written as

$$\begin{aligned} C_p &= C_{p,\mathrm{min}} + \left(C_{p,0} - C_{p,\mathrm{min}}\right)\cos^2\left(\frac{\pi}{2}\frac{\theta}{\theta_{\mathrm{min}}}\right) && 0^{\circ} \leqslant \theta \leqslant \theta_{\mathrm{min}} \\ C_p &= C_{p,\mathrm{min}} + \Delta C_{p,r}\cos^2\left(\frac{\pi}{2}\frac{\theta - \theta_{\mathrm{b}}}{\theta_{\mathrm{min}} - \theta_{\mathrm{b}}}\right) && \theta_{\mathrm{min}} < \theta \leqslant \theta_{\mathrm{b}} \\ C_p &= C_{p,\mathrm{b}} && \theta_{\mathrm{b}} < \theta \leqslant 180^{\circ} \end{aligned} \tag{2.11}$$

The angular dependence in (2.11) is consistent with the similarity solution developed by Yeung (2007) for incompressible flow over spheres. Reynolds number dependent functions for $C_{f,\mathrm{max}}$, $C_{p,\mathrm{min}}$ and $\Delta C_{p,r}$ will be determined in § 4. The incompressible expressions for $C_f$ and $C_p$ will be extended to compressible flow in the following subsections.

### 2.2. *Subsonic flow* ($0.3 \leqslant M_\infty < 1$)

As the freestream Mach number ($M_\infty$) increases and density variations within the flow become non-negligible, the scaling of $C_f$ and $C_p$ are influenced by changes the boundary layer properties. Following the preceding analysis for $C_f$, compressibility changes will be accounted for by modifying the expressions for $U_e$, $\rho_e$, and $\delta/R$ in the momentum-integral equation. The incompressible potential flow solution for $U_e$ can be extended to compressible flow through the Janzen-Rayleigh expansion, JRE (Janzen 1913; Rayleigh 1916). Kaplan (1940) used this expansion to compute $U_e$ for a sphere in compressible flow, which can be expressed as

$$\frac{U_e}{U_\infty} = u_0(\theta)\left[1 + u_1(\theta)M_\infty^2 + u_2(\theta)M_\infty^4 + ...\right] \tag{2.12}$$

where $u_0 = 3/2\sin\theta$, $u_1 = -0.2515 + 0.4602\sin^2\theta$ and $u_2 = -0.01707 - 0.5447\sin^2\theta + 0.7988\sin^4\theta$. Isentropic relations can be used to relate $\rho_e$ and $U_e$ along the edge of the boundary layer, resulting in the following expanded expression

$$\frac{\rho_e}{\rho_\infty} = \left[1 + \rho_1(\theta)M_\infty^2 + \rho_2(\theta)M_\infty^4 + \rho_3(\theta)M_\infty^6 + ...\right]^{1/\gamma-1} \tag{2.13}$$

where $\rho_1 = (\gamma-1)(1-u_0^2)/2$, $\rho_2 = -(\gamma-1)u_0^2 u_1$, $\rho_3 = -(\gamma-1)u_0^2(u_1^2 + 2u_2)/2$, and $\gamma$ is the ratio of specific heat capacities. It was shown in the previous subsection that $C_{f,\max}$ is primarily dependent on the second and third terms in (2.2) for the forward portion of the sphere. Since both terms reduce to approximately the same scaling dependence, $C_f$ will be estimated using only the second term, which represents the acceleration of the external flow. After substituting (2.12) and (2.13) into (2.2), and evaluating $\delta/R$ using $U_e$, $\rho_e$, and a reference temperature based power-law function for viscosity ($\mu^* \propto T^{*\omega}$), the scaling of $C_f$ in compressible flow becomes

$$C_f \sim C_f^{ic}\left(\frac{T^*}{T_\infty}\right)^{\omega/2}\left[\frac{1 + u_1 M_\infty^2 + u_2 M_\infty^4}{\left(1 + \rho_1 M_\infty^2 + \rho_2 M_\infty^4\right)^{1/1-\gamma}}\right]^{1/2}\left[1 + F_1(\theta)M_\infty^2 + F_2(\theta)M_\infty^4\right] \tag{2.14}$$

where $C_f^{ic}$ is equivalent to the incompressible skin friction scaling, $T^*/T_\infty$ is the reference temperature ratio, $F_1 = u_1 + u_1'\tan\theta$, and $F_2 = u_2 + u_2'\tan\theta$. Evaluating the third and fourth terms in (2.14) over the forebody of the sphere shows that the third term remains close to unity, and the primary Mach-number dependence for $C_f$ is from $1 + F_1(\theta)M_\infty^2 + F_2(\theta)M_\infty^4$, which comes directly from the velocity gradient ($dU_e/d\theta$). Therefore, an expression for $C_{f,\max}$ in subsonic flow can be approximated as

$$C_{f,\max} = C_{f,\max}^{ic}(Re_\infty)\left(\frac{T^*}{T_\infty}\right)^{\omega/2} F_M(M_\infty, \theta_{\max}) \tag{2.15}$$

where $F_M = 1 + F_1(\theta_{\max})M_\infty^2 + F_2(\theta_{\max})M_\infty^4$ and $C_{f,\max}^{ic}(Re_\infty)$ corresponds to the incompressible expression evaluated with $Re_\infty$. Given that the maximum shear stress typically occurs near the middle of the forebody where $F_1$ and $F_2$ are both positive, $C_{f,\max}$ will increase with $M_\infty$. Additionally, $C_{f,\max}$ is dependent on the surface temperature ($T_w$) of the sphere in compressible flow, which is captured through $T^*/T_\infty$. An expression for $T^*/T_\infty$ is given by Meador & Smart (2005) for compressible flow over flat plates, which can be rewritten in terms of the stagnation temperature ($T_0$) with

$$\frac{T^*}{T_\infty} = b_1 - b_2 r + \left[(1-b_1)\frac{T_w}{T_0} + b_2 r\right]\left[1 + \frac{\gamma-1}{2}M_\infty^2\right] \tag{2.16}$$

where $b_1$ and $b_2$ are unknown coefficients for spheres, and $r$ is the adiabatic wall recovery factor ($r = \sqrt{Pr}$). The values of $b_1$ and $b_2$ will be determined in § 2.4.

For isentropic flow at the boundary layer edge, the surface pressure can be related to $\rho_e$ through $p_w \propto \rho_e^\gamma$; therefore, an expression for $C_p$ can be written as

$$C_p = \frac{2}{\gamma M_\infty^2}\left[\left(\frac{\rho_e}{\rho_\infty}\right)^\gamma - 1\right] \tag{2.17}$$

It was shown in the previous subsection that compressibility effects contribute to the stagnation pressure through the inviscid term $C_{p,0,M}$ in (2.9). By evaluating (2.17) at the stagnation point ($U_e = 0$), an expression for $C_{p,0,M}$ in subsonic flow becomes

$$C_{p,0,M} = \frac{2}{\gamma M_\infty^2}\left[\left(1 + \frac{\gamma-1}{2}M_\infty^2\right)^{\frac{\gamma}{\gamma-1}} - 1\right] \tag{2.18}$$

By substituting (2.13) into (2.17), simplifying using a binomial expansion, and retaining terms up to $M_\infty^4$, $C_p$ can be approximated as a JRE correction to the incompressible result ($C_p^{ic}$) with

$$C_p \approx C_p^{ic}\left[1 + \left(\frac{\rho_2}{\rho_1} + \frac{\rho_1}{2(\gamma-1)}\right)M_\infty^2 + \left(\frac{\rho_3}{\rho_1} + \frac{\rho_2}{\gamma-1} + \frac{2-\gamma}{6(\gamma-1)^2}\rho_1^2\right)M_\infty^4\right] \tag{2.19}$$

where $C_p^{ic} = 1 - u_0^2$. Upon evaluation of the leading coefficients for the $M_\infty^2$ and $M_\infty^4$ terms in (2.19), both are found to be positive near the shoulder of the sphere as the flow approaches $\theta_{\min}$. Therefore, an expression for $C_{p,\min}$ in subsonic flow can be written as

$$C_{p,\min} = C_{p,\min}^{ic}(Re_\infty)P_M(M_\infty, \theta_{\min}) \tag{2.20}$$

where $P_M = 1 + P_1(\theta_{\min})M_\infty^2 + P_2(\theta_{\min})M_\infty^4$. This expression has the same low-Mach number expansion as Prandtl-Glauert transformation ($C_p^{ic}/\sqrt{1-M_\infty^2}$), which is used to relate the incompressible and compressible potential flow solutions for slender bodies. Similar to the analysis for incompressible flow, the scaling of $C_{p,\mathrm{b}}$ can be approximated using a pressure recovery relative to $C_{p,\min}$. By applying a similar JRE-type correction to the incompressible scaling ($\Delta C_{p,r}^{ic}$), an expression for $\Delta C_{p,r}$ in compressible flow can be approximated as

$$\Delta C_{p,r} = \Delta C_{p,r}^{ic}(Re_\infty)R_M(M_\infty, \theta_s) \tag{2.21}$$

where $R_M = 1 + R_1(\theta_s)M_\infty^2 + R_2(\theta_s)M_\infty^4$. When the freestream Mach number exceeds approximately 0.6, the flow near the shoulder of the sphere can accelerate to the speed of sound, which leads to the formation of a lambda shock wave. This shock wave interacts with the boundary layer and modifies the edge conditions ($U_e$, $\rho_e$, $p_e$). Because the JRE model was derived for subsonic flow and does not capture transonic effects, the influence of the lambda shock will be absorbed into the coefficients for $F_M$, $P_M$, and $R_M$, which will be determined in § 4.

### 2.3. *Supersonic flow* ($1 \leqslant M_\infty < 5$)

In supersonic flow, the detached bow shock rapidly compresses the gas in front of the sphere. Due to the large jump in density, pressure, and temperature across the shock, the reference flow variables should be approximated using the post-shock conditions ($Re_\infty \to Re_s$, $M_\infty \to M_s$). The normal-shock conditions (Rankine 1870; Hugoniot 1887) are used to compute the new post-shock reference variables which are given by

$$Re_s = Re_\infty \left[1 + \frac{2(\gamma-1)}{(\gamma+1)^2 M_\infty^2}\left(M_\infty^2 - 1\right)\left(1+\gamma M_\infty^2\right)\right]^{-\omega} \tag{2.22}$$

$$M_s = \left(\frac{(\gamma-1)M_\infty^2 + 2}{2\gamma M_\infty^2 - (\gamma-1)}\right)^{1/2} \tag{2.23}$$

After the flow is rapidly compressed by the shock, it expands and accelerates around the forebody of the sphere. Because the flow is confined to a finite region between the shock and the surface of the sphere, a thinner shock layer causes the flow to turn around the sphere over a very small distance, resulting in a stronger pressure gradient (and therefore rate of acceleration). This length scale corresponds to the shock standoff distance ($\Delta_s$), which is inversely proportional to the velocity gradient at the stagnation point ($dU_e/d\theta \propto R/\Delta_s$). Since the subsonic post-shock flow at the boundary layer edge can still be approximated using the JRE method, (2.2) can simply be evaluated using $U_e/U_s$ and $\rho_e/\rho_s$. To incorporate the effect of the shock layer thickness into the boundary layer scaling, $R/\Delta_s$ is included as a multiplier to the velocity gradient in (2.2), with the angular dependence being absorbed into the JRE coefficients. Furthermore, to keep the definition of $C_f$ consistent with (1.1), $C_{f,\max}$ must be normalized using the ratio of dynamic pressures across the shock ($q_s/q_\infty$), which reduces to $\rho_\infty/\rho_s$ through mass conservation. Therefore, an expression for $C_{f,\max}$ in supersonic flow can be written as

$$C_{f,\max} = C_{f,\max}^{ic}(Re_s)\left(\frac{T^*}{T_s}\right)^{\omega/2} F_M(M_s, \theta_{\max})\left(\frac{\rho_\infty}{\rho_s}\right)\left(\frac{R}{\Delta_s}\right) \tag{2.24}$$

where $F_M = F_{0,s} + F_{1,s}M_s^2 + F_{2,s}M_s^4$, and $T^*/T_s$ is equivalent to (2.16) evaluated with $M_s$.

Given that the viscosity correction factor in (2.9) was derived based on the stagnation point velocity gradient and displacement thickness referenced to the freestream conditions, the expression for the stagnation pressure in supersonic flow becomes

$$C_{p,0} = C_{p,0,M} + \left(\frac{12}{Re_s + 0.644\sqrt{Re_s}}\right)\left(\frac{\rho_\infty}{\rho_s}\right)\left(\frac{R}{\Delta_s}\right) \tag{2.25}$$

where

$$C_{p,0,M} = \frac{2}{\gamma M_\infty^2}\left[\left(\frac{2\gamma M_\infty^2 - (\gamma-1)}{\gamma+1}\right)\left(1 + \frac{\gamma-1}{2}M_s^2\right)^{\frac{\gamma}{\gamma-1}} - 1\right] \tag{2.26}$$

where $C_{p,0,M}$ is derived by first applying the normal-shock conditions to determine the pressure change across the shock, and then isentropically compressing the post-shock gas to the stagnation point. Since the shock standoff term ($R/\Delta_s$) is approximately equal to the density ratio across the shock $\rho_s/\rho_\infty$ (Inouye 1965), the combination of the last two terms in (2.24) and (2.25) reduce to approximately order-unity. The presence of the shock also shifts the location of $\theta_{\min}$ downstream of the shoulder and into the afterbody region, as the high-pressure post-shock flow continues to expand around the sphere. The flow in the afterbody region involves complex physical phenomena, including expansion/recompression waves, flow separation, vortex shedding, and shock-boundary-layer interactions, which are challenging to model in the supersonic regime. This region often requires the use of CFD solvers with turbulence models or direct numerical simulation (DNS) methods to resolve the flow. Therefore, $C_{p,\min}$ and $\Delta C_{p,r}$ (and therefore $C_{p,\mathrm{b}}$) are approximated using the scaling $C_p \sim -1/M_\infty^2$, which is commonly employed for the afterbody pressure distribution of spheres (Jernell 1970). To ensure the subsonic results ((2.20) and (2.21)) are maintained for $M_\infty = 1$, the expressions for $C_{p,\min}$ and $\Delta C_{p,r}$ in supersonic flow can be written as

$$C_{p,\min} = C^{ic}_{p,\min}(Re_\infty)\frac{P_M(M_\infty = 1)}{M_\infty^2} \tag{2.27a}$$

$$\Delta C_{p,r} = \Delta C^{ic}_{p,r}(Re_\infty)\frac{R_M(M_\infty = 1)}{M_\infty^2} \tag{2.27b}$$

This choice of scaling also ensures consistency in the hypersonic limit, which will be discussed in the next subsection.

### 2.4. *Scaling in the hypersonic limit ($M_\infty \geqslant 5$)*

In hypersonic flows over blunt bodies, a significant fraction of the freestream kinetic energy is converted to internal energy modes across the shock wave. This results in a high-temperature, partially dissociated post-shock gas that is in thermal and chemical nonequilibrium. Additionally, the dissociated gas can react with the surface, altering the heat flux and species concentrations within the boundary layer. Because the physics involved in hypersonic flows are very challenging to model and often result in empirical approximations, the objective of this subsection will be to bridge the supersonic expressions for $C_f$ and $C_p$ with known values in the hypersonic limit. Approximations for $C_{f,\max}$ have been obtained based on the simulations of Golovachov (1985), which include hypersonic flow over a cooled ($T_w/T_0 = 0.15$) and adiabatic ($T_w = T_{aw}$) sphere at various $Re_s$. The limiting values of $C_{f,\max}$ are approximated as $1.84Re_s^{-0.5}$ and $2.33Re_s^{-0.5}$ for the cooled and adiabatic sphere boundary conditions, respectively. Due to the significant temperature gradients across the boundary layer in hypersonic flow, the coefficients in the reference temperature model ($b_1$ and $b_2$) should be constrained to data in this regime. This can be achieved by substituting the known expressions for $C^{ic}_{f,\max}$ and $M_s$ at high $M_\infty$ into (2.24), and evaluating $T_w/T_0$ at the corresponding boundary conditions. For the adiabatic wall condition, the surface temperature ratio is computed with

$$\frac{T_{aw}}{T_0} = \frac{1 + r\frac{\gamma-1}{2}M_s^2}{1 + \frac{\gamma-1}{2}M_s^2} \tag{2.28}$$

After equating the resulting expressions for $C_{f,\max}$ to the limiting values listed above, $b_1$ and $b_2$ can be written as

$$b_1 = 1 - \frac{0.099}{F_\infty^4} \tag{2.29a}$$

$$b_2 = \frac{5.723}{F_\infty^4} - 41.208 \tag{2.29b}$$

where $F_\infty$ is a constant that corresponds to $F_M(M_\infty \gg 1)$. In hypersonic flow, Paredes, Choudhari & Li (2017) found that $\tau_w$ peaks approximately 5° downstream of the sonic point on a spherical forebody, $\theta_{\mathrm{max}} \approx \theta^* + 5^{\mathrm{o}}$, where the angular location of $\theta^*$ is given by Lunev (2009) as

$$\theta^* = 34^{\mathrm{o}} + 40^{\mathrm{o}} \left( \frac{\gamma - 1}{\gamma + 1} \right) \tag{2.30}$$

where $\theta^* = 41^{\mathrm{o}}$ for $M_\infty \gg 1$ ($\gamma = 1.4$).

The pressure distribution in hypersonic flow can be modeled using modified Newtonian aerodynamics. In the original Newtonian aerodynamics approach, the surface pressure is approximated as the change in momentum normal to the surface times the surface normal mass flux (Newton 1687). After normalizing by the freestream dynamic pressure, and converting the surface inclination angle to $\theta$ (measured from the stagnation point), the expression for $C_p$ becomes

$$C_p = C_{p,0} \cos^2 \theta \tag{2.31}$$

where $C_{p,0} = 2$ for the original approach. The modified version involves computing the stagnation pressure coefficient as the value of (2.26) taken in the limit of infinite $Re_\infty$ and $M_\infty$ ($C_{p,0} = 1.839$ for $\gamma = 1.4$). Newtonian aerodynamics only applies (2.31) to forebody surfaces which have normal vectors facing the incoming flow. The afterbody region is typically modeled using $C_p = 0$, which is consistent with (2.27) evaluated for $M_\infty \gg 1$. Additionally, to ensure consistency between (2.11) and (2.31), $\theta_{\mathrm{min}}$ and $\theta_{\mathrm{b}}$ must equal $90^{\mathrm{o}}$ and $180^{\mathrm{o}}$ in the hypersonic limit, respectively.

## 3. Rarefied regime

### 3.1. *Velocity slip effects* ($0.01 \leqslant Kn_\infty < 0.1$)

In the slip regime, an extension to the Navier-Stokes equations can be derived using the first-order Chapman-Enskog solution of the Boltzmann equation (Chapman & Cowling 1970). The surface boundary conditions that are consistent with this derivation include the velocity slip and temperature jump conditions. Maxwell (1879) derived an expression for the velocity slip boundary condition by considering the tangential momentum exchange between particles and the surface. Assuming that a fraction of the incoming particles diffusely reflect $\sigma$, while the other fraction $(1 - \sigma)$ reflect specularly and retain their incoming tangential momentum, the slip velocity can be written as

$$u_{slip} = \frac{2 - \sigma}{\sigma} \lambda \left. \frac{\partial u}{\partial n} \right|_w \tag{3.1}$$

where $\lambda$ is the mean free path, and $\partial u / \partial n|_w$ is the velocity gradient normal to the surface. Like the solution of the Navier-Stokes equations in the continuum regime for low $Re_\infty$, analytical expressions for the shear stress and surface pressure in the slip regime can be derived using (3.1) as a surface boundary condition. Barber and Emerson (2000) carried out this derivation and determined expressions for $\tau_{r\theta}$, $\tau_{rr}$, and $p_w$ which correspond to the shear stress, normal stress, and surface pressure, respectively. Since the slip condition creates a nonzero radial velocity gradient at the surface, the additional viscous normal stress term contributes to the surface stress tensor, so $C_p$ must now be defined using the total normal stress acting on the surface $(-p_w + \tau_{rr})$. Therefore, low $Re_\infty$ expressions for $C_f$ and $C_p$ in the slip regime can be computed as

$$C_f = C_f^c \left( \frac{1}{1 + 6c_s Kn_\infty} \right) \tag{3.2a}$$

$$C_p = C_p^c \left( \frac{1 + 12c_s Kn_\infty}{1 + 6c_s Kn_\infty} \right) \tag{3.2b}$$

where $C_f^c$ and $C_p^c$ correspond to the continuum expressions, which are given by (2.1) and (2.7a), respectively, and $c_s = (2 - \sigma)/\sigma$.

As $Re_\infty$ increases and the boundary layer flow develops over the surface, the scaling of the surface shear stress can be determined by approximating the velocity gradient as $\partial u/\partial n|_w \sim (U_e - u_{slip})/\delta_s$. After substituting this expression into $\tau_w = \mu \, \partial u/\partial n|_w$ and applying the slip condition (3.1), $C_f$ can be approximated as

$$C_f = C_f^c \left( \frac{1}{1 + C_1 c_s Kn_\infty Re_\infty^{0.5}} \right) \tag{3.3}$$

where $C_1$ is an order-unity coefficient that incorporates the shape of the slip velocity profile as well as the scaling between the no-slip and slip boundary layer thicknesses. The term $c_s Kn_\infty Re_\infty^{0.5}$ represents the ratio of the slip length and boundary layer thickness which is consistent with the nondimensional slip parameter derived by Martin & Boyd (2006) for laminar flow over a flat plate. Since the viscous normal stress and surface pressure are both primarily governed by streamwise velocity gradients along the surface, the diameter of the sphere is still the relevant length scale as $Re_\infty$ increases. Therefore, the scaling of the slip coefficients remains consistent with the low $Re_\infty$ result. Slip correction factors for $C_f$ and $C_p$ can now be approximated as

$$\beta_\tau = \frac{1}{1 + 6c_s Kn_\infty [1 + (C_1/6)^2 Re_\infty]^{0.5}} \tag{3.4a}$$

$$\beta_p = \frac{1 + 12c_s Kn_\infty}{1 + 6c_s Kn_\infty} \tag{3.4b}$$

where $\beta_\tau$ and $\beta_p$ correspond to $C_f = C_f^c \beta_\tau$ and $C_p = C_p^c \beta_p$, respectively. The slip correction factors reduce to unity in the continuum limit ($Kn_\infty \to 0$) and are valid for low-speed flows. An extension to $\beta_\tau$ and $\beta_p$ in the free-molecular limit will be derived in the following subsections.

### 3.2. *Analytical expressions for free-molecular flow ($Kn_\infty \geqslant 10$)*

In the free-molecular regime, analytical expressions for $C_f$ and $C_p$ can be derived by considering the momentum and energy exchange with the surface due to the incident and reemitted particles. For spheres, the expressions for $C_f$ and $C_p$ are given by Schaaf & Chambre (1958) as

$$C_f^{fm} = \frac{\sigma_t \sin\theta}{s\sqrt{\pi}} \left[ e^{-\xi^2} + \xi\sqrt{\pi}\,(1 + \operatorname{erf}(\xi)) \right] \tag{3.5a}$$

$$C_p^{fm} = \frac{1}{s^2} \left[ \left( \frac{2 - \sigma_n}{\sqrt{\pi}} \xi + \frac{\sigma_n}{2} \sqrt{\frac{T_w}{T_\infty}} \right) e^{-\xi^2} + \left\{ (2 - \sigma_n) \left( \xi^2 + \frac{1}{2} \right) + \frac{\sigma_n}{2} \xi \sqrt{\pi} \sqrt{\frac{T_w}{T_\infty}} \right\} (1 + \operatorname{erf}(\xi)) - 1 \right] \tag{3.5b}$$

where the superscript "$fm$" denotes free-molecular flow, $s = M_\infty\sqrt{\gamma/2}$, and $\xi = s\cos\theta$. The accommodation coefficients $\sigma_t$ and $\sigma_n$ are defined as the fraction of particles that exchange momentum with the surface in the tangential and normal directions, respectively. Two hypothetical cases exist in which every particle either undergoes a specular reflection or a diffuse reflection. These cases correspond to $\sigma_t = \sigma_n = 0$ and $\sigma_t = \sigma_n = 1$, respectively. Materials with near specular surfaces experience significantly less drag compared to rough surfaces with many diffuse collisions.

### 3.3. *Rarefaction correction in the transition regime* ($0.1 \leqslant Kn_\infty < 10$)

In general, it is very challenging to obtain analytical solutions to the Boltzmann equation for $C_f$ and $C_p$ across the full transition regime. Therefore, we will attempt to bridge the slip and free-molecular results discussed in the previous subsections. For low-speed flow in the free-molecular limit ($Kn_\infty \gg 1$), the ratio of (3.5*a*) and $C_f^c\beta_\tau$ can be reduced to $c_s\sigma_t$. To maintain the correct scaling in both the continuum and free-molecular limits, the following rarefaction correction factor is proposed

$$\phi_\tau = \frac{1 + c_s\sigma_t Kn_\infty}{1 + Kn_\infty} \tag{3.6}$$

where $C_f = C_f^c\beta_\tau\phi_\tau$. Basically, $\phi_\tau$ reduces to unity for $Kn_\infty \ll 1$ and $c_s\sigma_t$ for $Kn_\infty \gg 1$. It is important to note that the slip coefficient ($c_s$) was kept generalized for this derivation, so any expression for $c_s$ can be implemented. For this work, Maxwell's slip coefficient is used with $\sigma = \sigma_t$. The same procedure can be used to determine the corresponding rarefaction correction factor for $C_p$ across the transition regime, resulting in

$$\phi_p = \frac{1}{1 + 3Kn_\infty(1 - 0.1073\sigma_n)^{-1}} \tag{3.7}$$

where $C_p = C_p^c\beta_p\phi_p$. If one computes the drag coefficient ($C_d$) by integrating the new expressions for $C_f$ and $C_p$ over the surface of the sphere, the result is approximately equal to the low-speed rarefaction correction proposed by Davies, $C_d = C_d^c f_{Kn}$ (Davies 1945).

To extend $\beta_\tau\phi_\tau$ and $\beta_p\phi_p$ to high-Mach-number flows, the high-speed rarefaction parameter $W_r^T$ is employed. For high-speed flow over spheres in the transition regime, $W_r^T$ was used as the scaling factor to extend $C_d^c f_{Kn}$ to high $M_\infty$ and $Kn_\infty$ flows (Singh *et al.* 2022), which makes it the appropriate scaling parameter for $C_f$ and $C_p$ in this work. Following the approach taken by Singh *et al.* (2022) to modify $f_{Kn}$, a general expression for the rarefaction correction factors can be written as

$$\psi_{Wr} = \frac{\beta\phi}{1 + 1.27W_r^T} \tag{3.8}$$

where

$$W_r^T = \frac{M_\infty^{2\omega}}{Re_\infty}\left(1 + \frac{T_w}{T_0}\right)^\omega \tag{3.9}$$

Since the scaling parameter $W_r^T$ was derived using second and third-order Chapman-Enskog solutions of the Boltzmann equation (Singh & Schwartzentruber 2016; 2017), it only remains accurate up to finite $Kn_\infty$. Therefore, to recover (3.5) in the free-molecular limit, $C_f^c\psi_{Wr,\tau}$ and $C_p^c\psi_{Wr,p}$ must be bridged to $C_f^{fm}$ and $C_p^{fm}$, respectively. Following the approach taken by Singh *et al.* (2022) to bridge the continuum and free-molecular drag coefficients, the final expressions for $C_f$ and $C_p$ can be written as

$$C_f = C_f^c \frac{\psi_{W_r,\tau}}{1+Br^\eta} + C_f^{fm} \frac{Br^\eta}{1+Br^\eta} \tag{3.10a}$$

$$C_p = C_p^c \frac{\psi_{W_r,p}}{1+Br^\eta} + C_p^{fm} \frac{Br^\eta}{1+Br^\eta} \tag{3.10b}$$

where $\eta = 1.8$ and

$$Br = W_r^T \frac{M_\infty^{2\omega-1}+1}{M_\infty^{2\omega-1}} \tag{3.11}$$

Therefore, given the current flow regime of the sphere relative to its surrounding gas ($Re_\infty$, $M_\infty$, $Kn_\infty$), local gas properties ($\gamma$, $\omega$, $Pr$), and surface conditions ($T_w/T_0$, $\sigma_t$, $\sigma_n$), the distributions of the local shear stress and surface pressure around the sphere can be determined. In the following section, the unknown incompressible parameters ($C_{f,\max}^{ic}$, $C_{p,\min}^{ic}$, $\Delta C_{p,r}^{ic}$) and the corresponding JRE-based functions ($F_M$, $P_M$, $R_M$) are determined based on simulated and experimental data in the continuum regime. Finally, for ease of implementation, the complete set of model equations, including the angular positions, are provided in the Appendix.

## 4. Estimation of continuum parameters

In the continuum regime, it was shown that the incompressible expressions for $C_f$ and $C_p$ are dependent on the unknown parameters $C_{f,\max}^{ic}$, $C_{p,\min}^{ic}$, and $\Delta C_{p,r}^{ic}$. It was determined that $C_{f,\max}^{ic}$ is equivalent to $6/Re_\infty$ in Stokes limit ($Re_\infty \ll 1$) and $3.8Re_\infty^{-0.5}$ prior to the critical transition condition of the boundary layer. The following expression was obtained based on simulated data at subcritical $Re_\infty$, while maintaining the correct limiting behavior

$$C_{f,\max}^{ic} = \frac{6}{Re_\infty}\left(1+\frac{Re_\infty}{2.493}\right)^{0.5} \tag{4.1}$$

A comparison between (4.1) and the values extracted from simulated data is shown in figure 1, with excellent agreement across the full subcritical regime. Similar to behavior of $C_{f,\max}^{ic}$ at low $Re_\infty$, it was shown that $C_{p,\min}^{ic}$ and $\Delta C_{p,r}^{ic}$ reduce to $-6/Re_\infty$ and 0 in the Stokes flow limit, respectively. Since the scaling of $C_{p,\min}^{ic}$ and $\Delta C_{p,r}^{ic}$ are strongly influenced by boundary layer separation and wake development at high $Re_\infty$, the expressions were anchored to the Reynolds numbers that correspond to the onset of separation ($Re_\infty = 20$) and wake unsteadiness ($Re_\infty = 270$). The following expressions for $C_{p,\min}^{ic}$ and $\Delta C_{p,r}^{ic}$ were obtained from simulated and experimental data across the subcritical regime

$$C_{p,\min}^{ic} = -\frac{6}{Re_\infty}\left(1+\frac{Re_\infty^{2.5}}{270}\right)^{0.384} \tag{4.2a}$$

$$\Delta C_{p,r}^{ic} = \frac{(Re_\infty/20)^{1.586}}{1+168.5(Re_\infty/270)^{1.72}} \tag{4.2b}$$

Figure 2 compares (4.2) with the corresponding simulated and experimental data, showing great agreement across the full subcritical flow regime.

In compressible flow, it was shown that $C_{f,\max}$, $C_{p,\min}$, and $\Delta C_{p,r}$ are proportional to the JRE-based functions $F_M$, $P_M$, and $R_M$, respectively. Additionally, it was shown that $F_M$, $P_M$, and $R_M$ are dependent on the freestream Mach number in subsonic flow and also the angular positions $\theta_{\max}$, $\theta_{\min}$, and $\theta_s$, respectively. However, to extend these functions to transonic flow, the angular dependence in the leading coefficients is neglected, and the coefficients are fit as constants using simulated data. Since $F_M$ is dependent on the post-shock Mach number in supersonic flow, it is fit as a piecewise expression. Based on simulated data of compressible flow over spheres, a piecewise expression for $F_M$ was obtained as

$$F_M = 1 + 0.143 M_\infty^2 - 0.243 M_\infty^4 \qquad M_\infty < 1 \tag{4.3a}$$

$$F_M = 0.56 + 0.351 M_s^2 - 0.011 M_s^4 \qquad M_\infty \geqslant 1 \tag{4.3b}$$

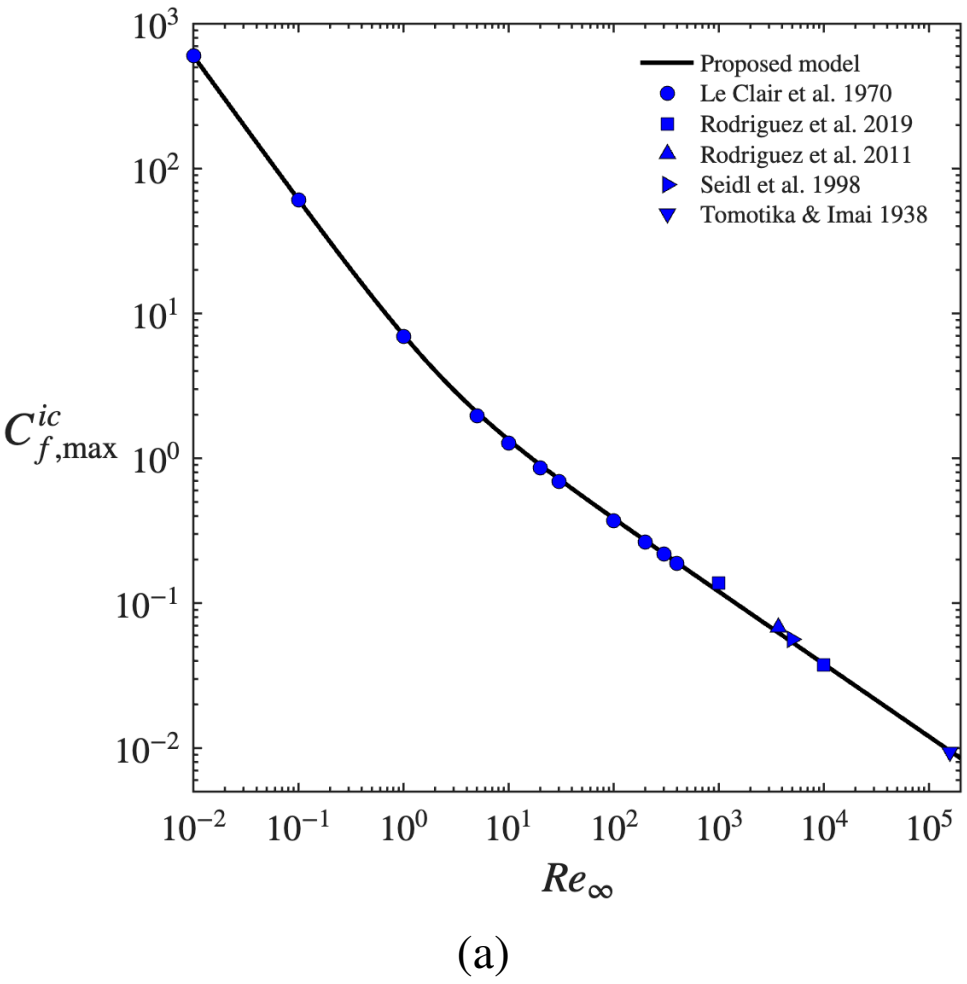

Figure 1: Comparison of $C^{ic}_{f,\max}$ with simulated data at subcritical $Re_\infty$.

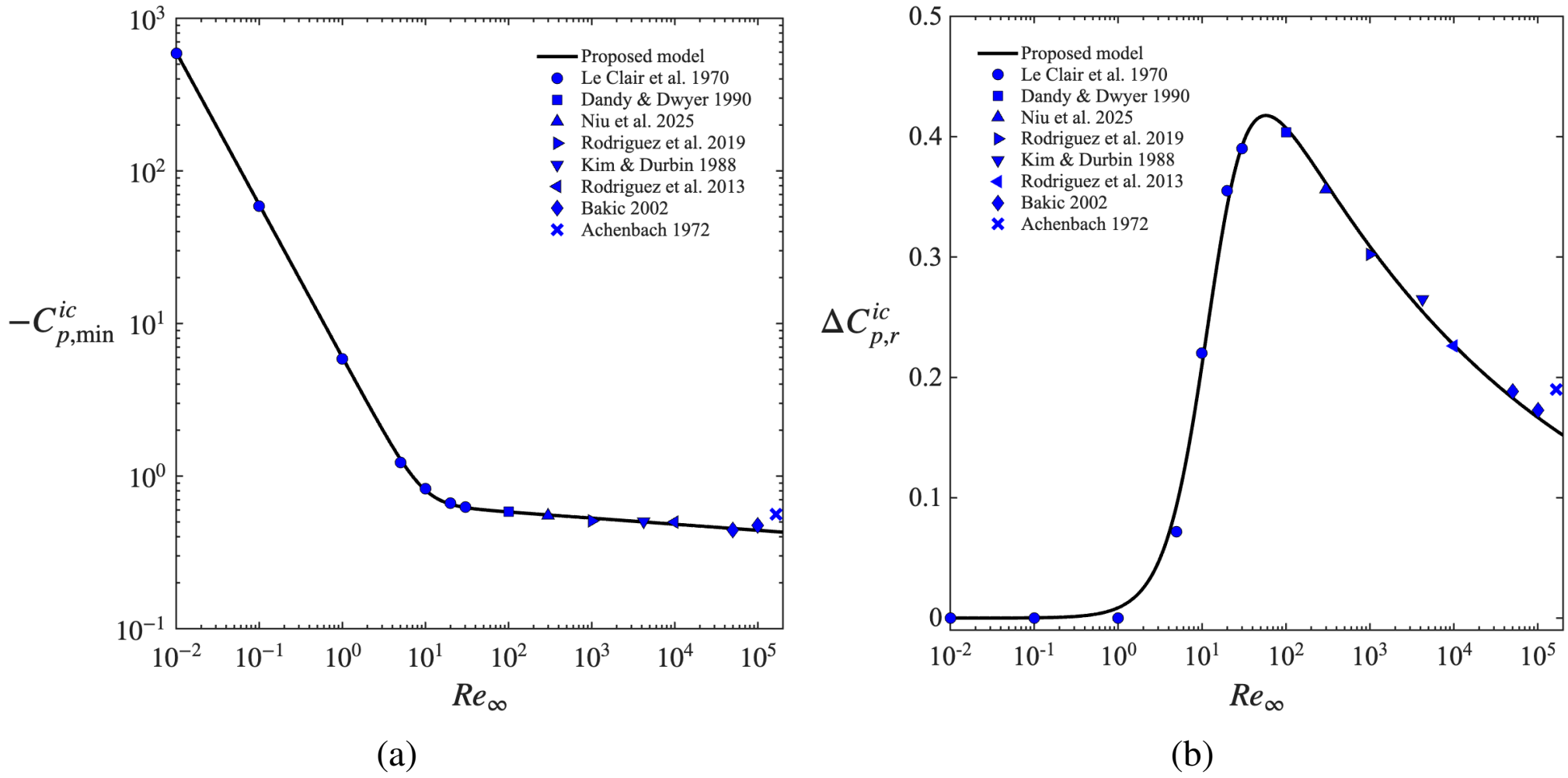

Figure 2: Comparison of (a) $C^{ic}_{p,\min}$ and (b) $\Delta C^{ic}_{p,r}$ with simulated and experimental data at subcritical $Re_\infty$.

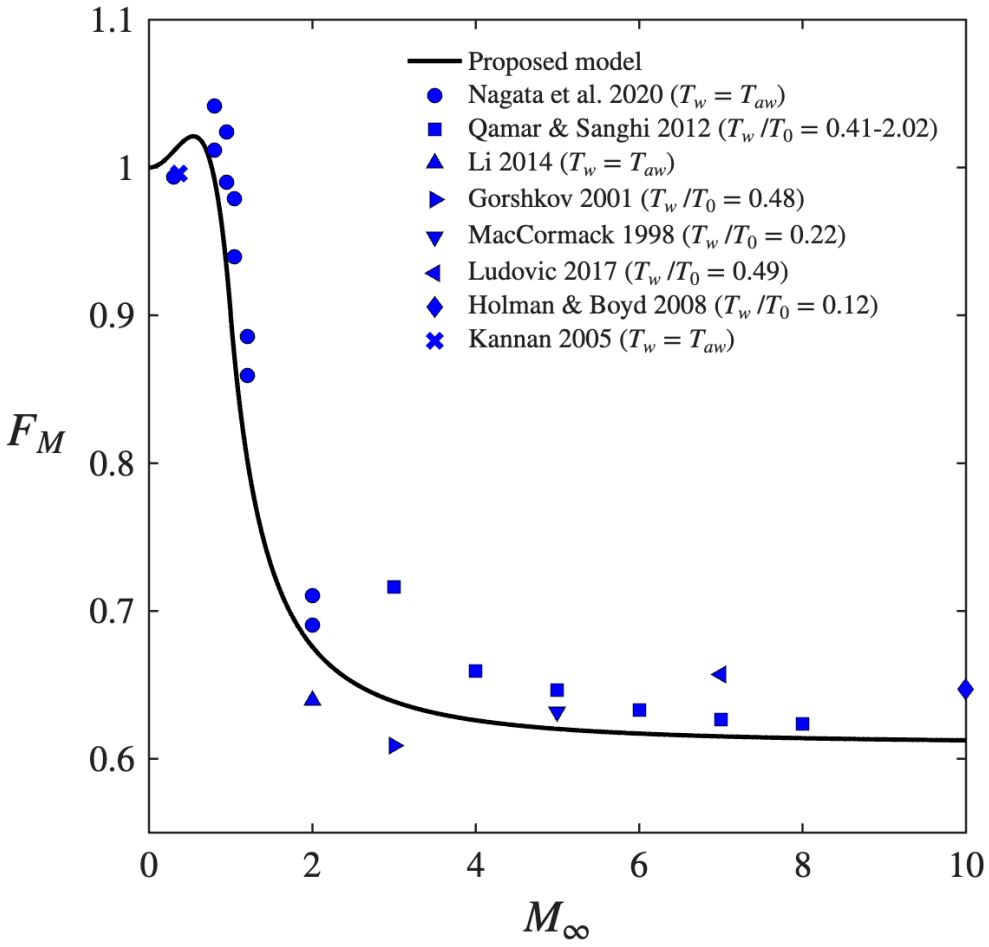

Figure 3: Comparison of $F_M$ with simulated data in compressible flow.

where $F_M = 0.61$ for $M_\infty \gg 1$. The reference temperature coefficients can now be computed using (2.29), which gives $b_1 = 0.285$ and $b_2 = 0.126$. These values are of the same order as the coefficients reported by Meador & Smart (2005) for laminar compressible flow over a flat plate ($b_1 = 0.45$ and $b_2 = 0.16$). A comparison between $F_M$ and the corresponding simulated data is shown in figure 3. In general, there is good agreement across the subsonic, supersonic, and hypersonic flow conditions. Based on simulated data of pressure distributions around spheres in subsonic flow, expressions for $P_M$ and $R_M$ were fit as

$$P_M = 1 + 0.174 M_\infty^2 + 0.045 M_\infty^4 \tag{4.4}$$

$$R_M = 1 + 0.230 M_\infty^2 - 0.452 M_\infty^4 \tag{4.5}$$

To recover the correct values of (4.4) and (4.5) for $M_\infty = 1$, the supersonic scaling is equivalent to $1.219/M_\infty^2$ and $0.778/M_\infty^2$ for $C_{p,\min}/C_{p,\min}^{ic}$ and $\Delta C_{p,r}/\Delta C_{p,r}^{ic}$, respectively. Because limited data is available in literature for the full pressure distributions around spheres, comparisons are restricted to subsonic and low supersonic flow conditions. A comparison between the simulated data and proposed expressions for $C_{p,\min}/C_{p,\min}^{ic}$ and $\Delta C_{p,r}/\Delta C_{p,r}^{ic}$ is shown in figure 4. Overall, there is good agreement between the proposed expressions and simulated data across these conditions.

## 5. Method accuracy in the continuum regime

In this section, the proposed models for $C_f$ and $C_p$ are compared with CFD simulations of incompressible and compressible flow over spheres in the continuum regime. The simulated cases were selected in order to validate the models ability to capture the relevant physical effects, including variations of $Re_\infty$, $M_\infty$, and $T_w/T_0$. The gas properties for the following incompressible and compressible cases correspond to air ($\gamma = 1.4, \omega = 0.75, Pr = 0.72$). The models are first compared to simulated results of incompressible flow across the subcritical regime. Figures 5 and 6 compare the $C_f$ and $C_p$ models with the corresponding simulated data, showing great agreement over a wide range of subcritical $Re_\infty$. The distributions of $C_f$ are captured very well by the angular scaling dependence that was determined from the momentum-integral equation. The acceleration of the flow dominates the distribution over the forebody of the sphere, whereas the rate of change of the momentum thickness dominates the distribution on the afterbody region. Similarly, the distributions of $C_p$ are captured very well by the stagnation, minimum, and base pressure coefficients. At low $Re_\infty$, the recovery pressure $\Delta C_{p,r}$ (and therefore $C_{p,\mathrm{b}}$) are strongly correlated to the length of the recirculation region, as the wake remains steady and axisymmetric. For higher $Re_\infty$, when the separation point stabilizes near the shoulder of the sphere, the wake becomes unstable and the distribution of $C_p$ remains nearly constant with increasing $Re_\infty$. This results in the plateau of the drag coefficient prior to the critical transition condition.

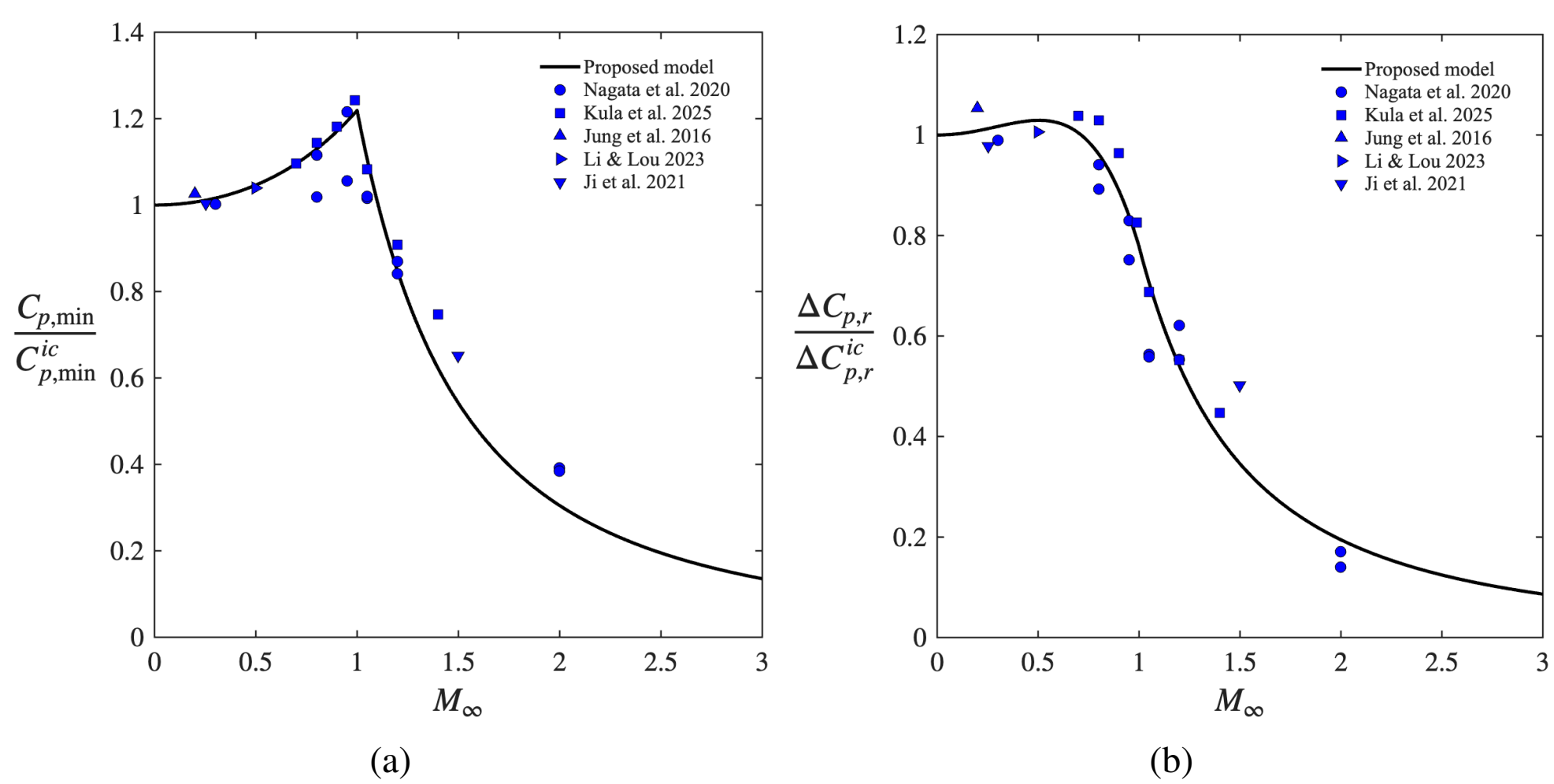

Figure 4: Comparison of (a) $C_{p,\min}/C^{ic}_{p,\min}$ and (b) $\Delta C_{p,r}/\Delta C^{ic}_{p,r}$ with simulated data in compressible flow.

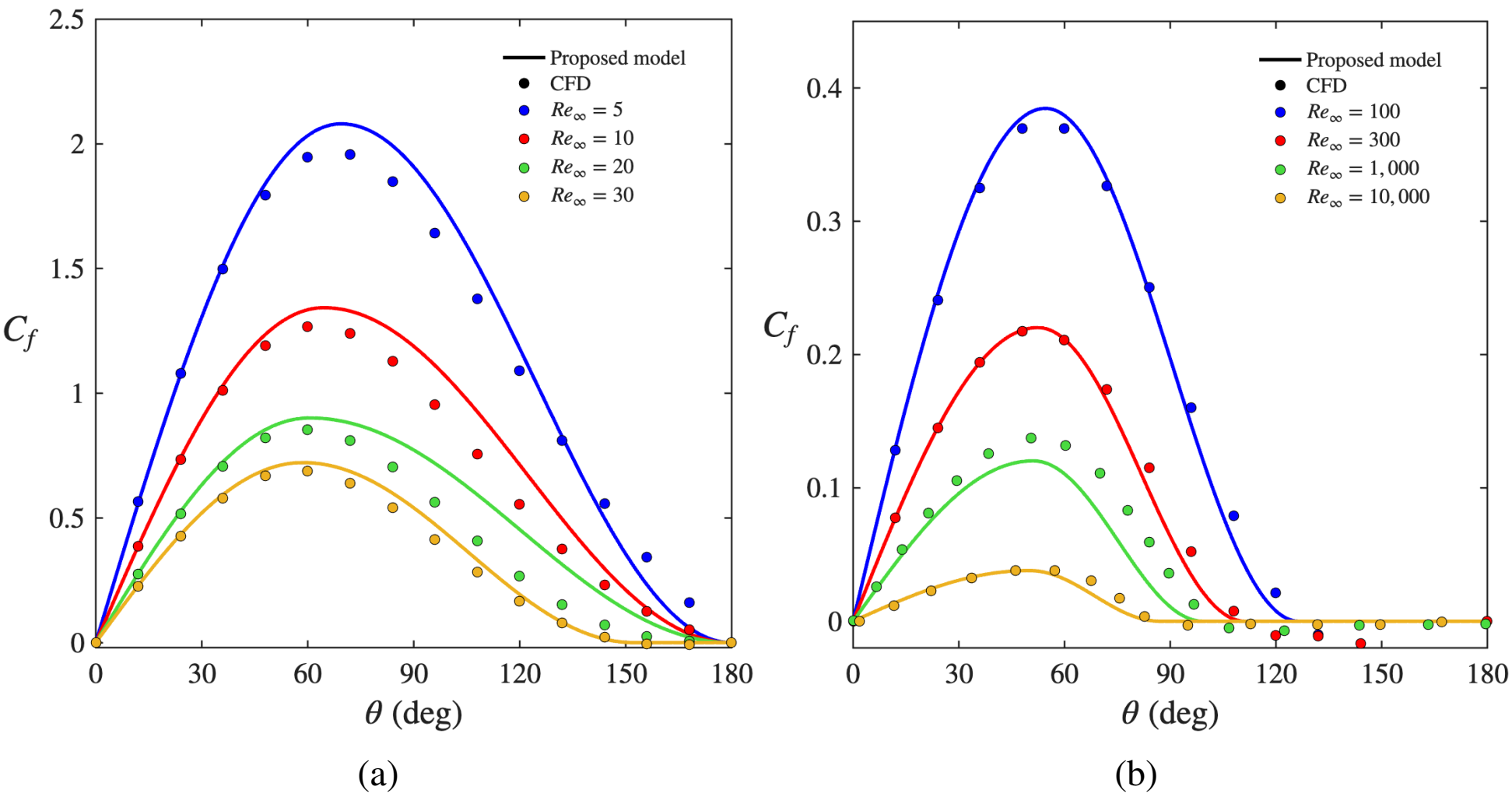

Figure 5: Comparison of $C_f$ distributions in incompressible flow at subcritical $Re_\infty$. CFD data: $Re_\infty = 5 - 300$ (Le Clair et al. 1970); $Re_\infty = 10^3, 10^4$ (Rodriguez et al. 2019).

Comparisons of the $C_f$ and $C_p$ distributions between the proposed model and CFD data in compressible flow are shown in figures 7 and 8, respectively. In general, the proposed model for $C_f$ shows good agreement with the simulated data over a wide range of $Re_\infty$, $M_\infty$, and $T_w/T_0$. The modification of the incompressible boundary layer properties using the JRE and reference temperature models is effective in capturing variations of $C_f$ with $M_\infty$ and $T_w/T_0$. Similarly, the proposed JRE-based functions for $C_p$ ($P_M$ and $R_M$) show good agreement with the simulated data over a range of $Re_\infty$ and $M_\infty$. The angular dependence of $C_f$ and $C_p$ is still captured well by the incompressible scaling.

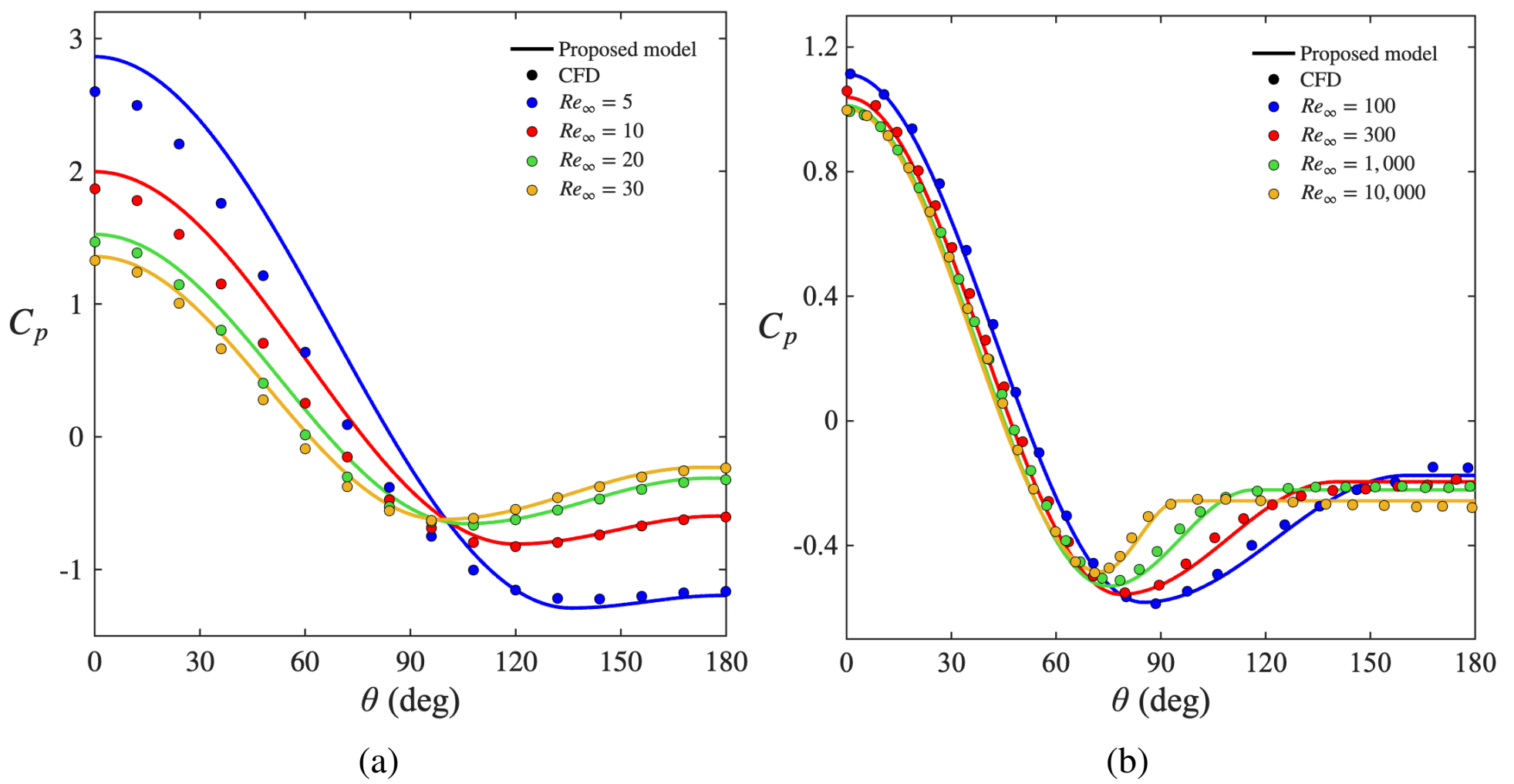

Figure 6: Comparison of $C_p$ distributions in incompressible flow at subcritical $Re_\infty$. CFD data: $Re_\infty = 5 - 30$ (Le Clair et al. 1970); $Re_\infty = 100$ (Dandy & Dwyer 1990); $Re_\infty = 300$ (Niu et al. 2025); $Re_\infty = 10^3$ (Rodriguez et al. 2019); $Re_\infty = 10^4$ (Rodriguez et al. 2013).

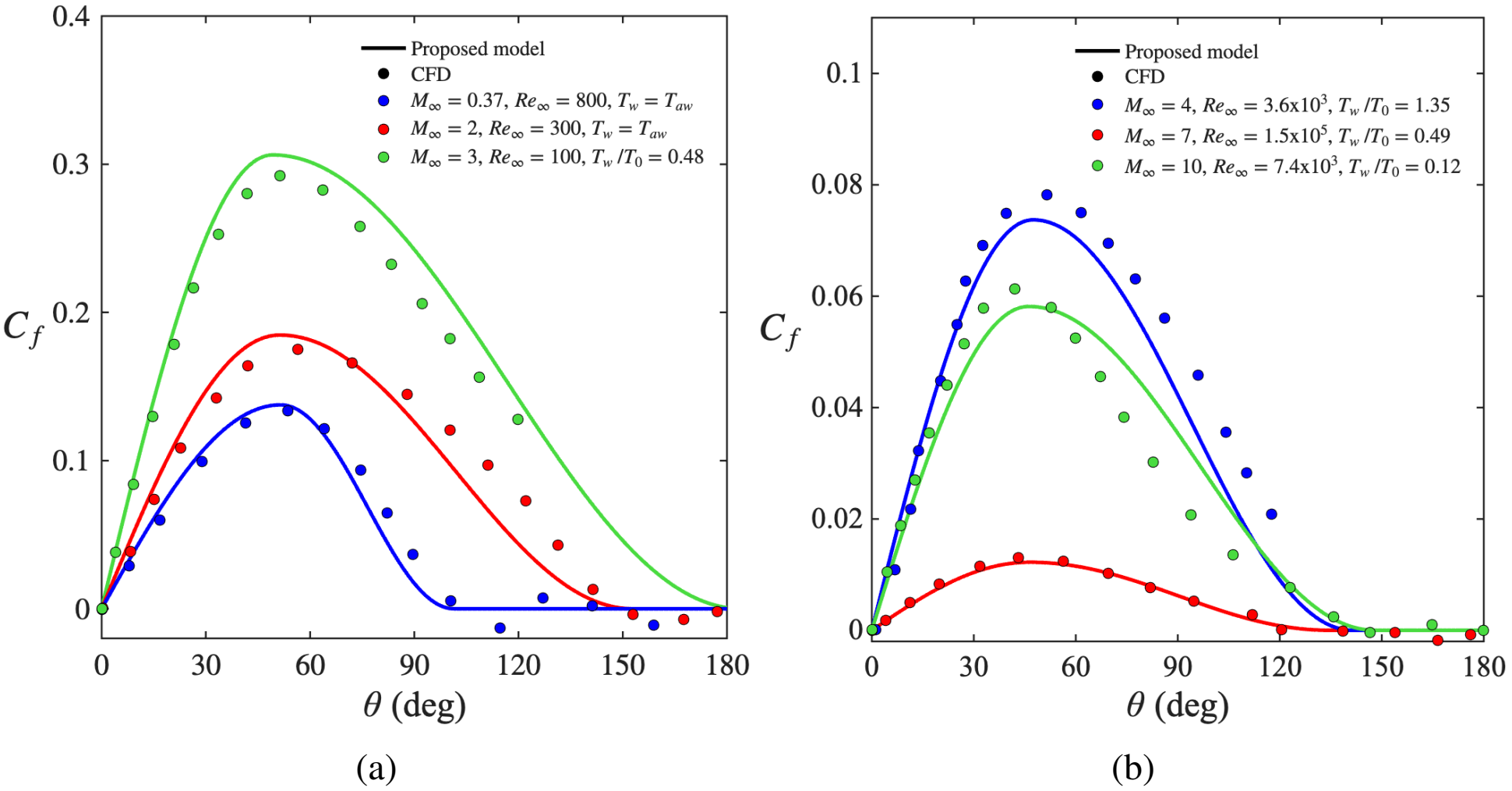

Figure 7: Comparison of $C_f$ distributions in compressible flow. CFD data: $M_\infty = 0.37$ (Kanna 2005); $M_\infty = 2$ (Li 2014); $M_\infty = 3$ (Gorshkov 2001, no-slip surface); $M_\infty = 4$ (Qamar & Sanghi 2012); $M_\infty = 7$ (Ludovic 2017); $M_\infty = 10$ (Holman & Boyd 2008).

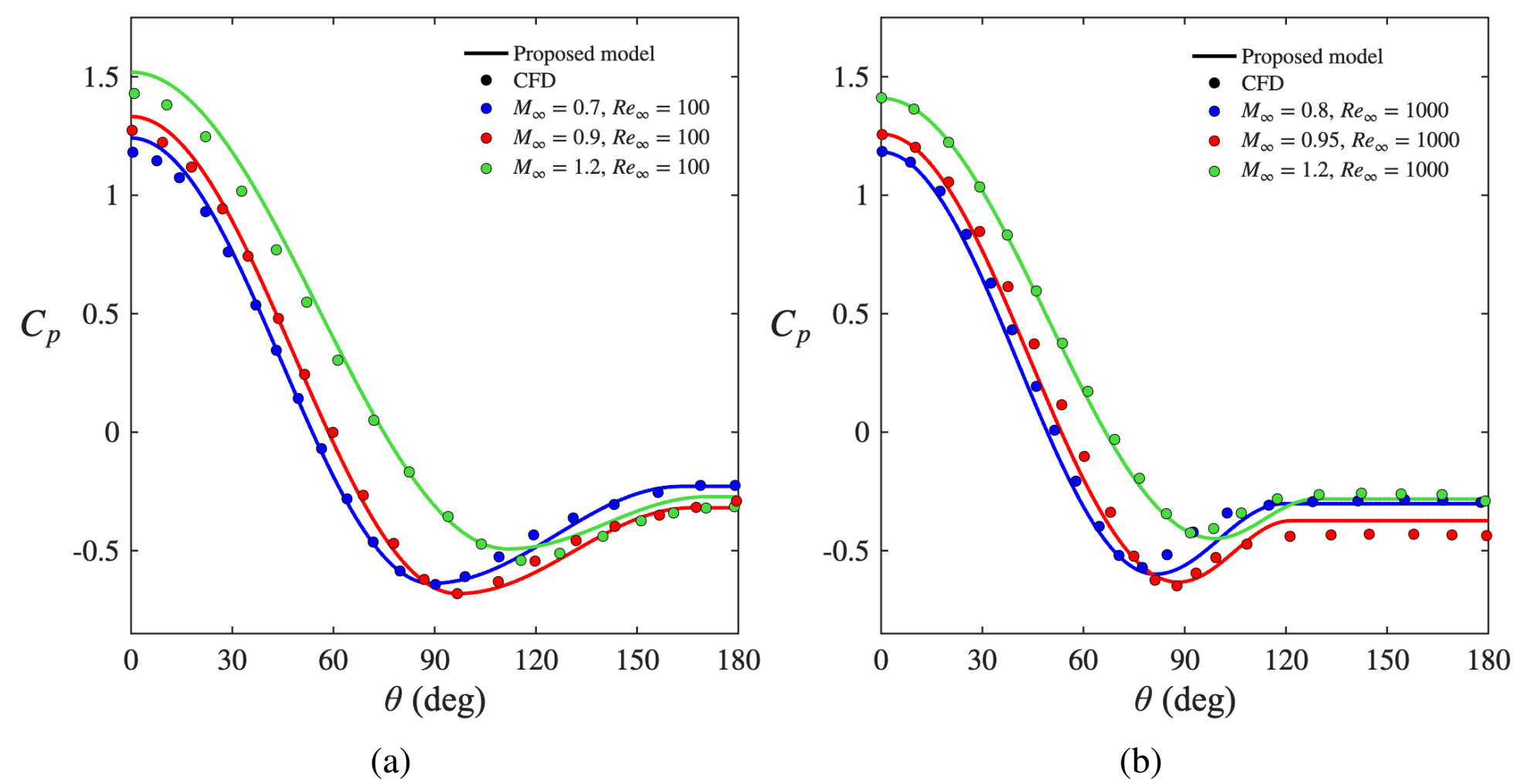

Figure 8: Comparison of $C_p$ distributions in compressible flow. CFD data: $Re_\infty = 100$ (Kula, Aslangil & Wong 2025); $Re_\infty = 1000$ (Nagata et al. 2020).

## 6. Method accuracy in the rarefied regime

In the rarefied regime, it was shown that velocity slip and gas-surface accommodation affect the distributions of $C_f$ and $C_p$. To extend $C_f$ and $C_p$ to high $Kn_\infty$ and $M_\infty$ flows, the high-speed rarefaction scaling parameter $W_r^T$ was chosen. Therefore, the goal of this section is to validate the proposed models ability to capture variations of $Kn_\infty$, $M_\infty$, and $\sigma_t/\sigma_n$. In a series of publications by Sharipov & Volkov (2022; 2024), DSMC simulations were performed on a sphere in a monatomic gas based on *ab initio* interatomic potentials at subsonic, supersonic, and hypersonic flow conditions. The surface of the sphere was modeled with the Cercignani-Lampis (CL) gas-surface interaction model (Cercignani & Lampis 1971), which contains two accommodation coefficients: the tangential momentum ($\alpha_t$) and normal energy ($\alpha_n$) coefficients. Comparisons of $C_f$ and $C_p$ in helium were made for the slip and transitional regimes with fully diffuse ($\alpha_t = \alpha_n = 1$) and partial ($\alpha_t = 0.4$ and $\alpha_n = 0.01$) surface accommodation at $M_\infty = 0.2$, 0.5, 1, and 2. Based on the simulated flow conditions, the parameters used in the proposed model include: $\gamma = 1.67$, $Pr = 0.67$, and $T_w = T_\infty = 300K$. Since the use of interatomic potentials does not directly translate to a single viscosity power-law form ($\mu \propto T^\omega$), the value of $\omega$ was chosen based on the variable hard-sphere (VHS) value for helium ($\omega = 0.66$).

For the following comparisons, the slip parameter $C_1$ was set equal to 2.08. The surface is modeled as fully diffuse for the first comparison, which corresponds to $\sigma_t = \sigma_n = 1$ in the proposed model. Comparisons of the $C_f$ and $C_p$ distributions between the proposed model and DSMC simulations are shown in figures 9 and 10, respectively. Overall, there is good agreement across a wide range of $Kn_\infty$ and $M_\infty$. The use of the slip model results in very good agreement for the low $Kn_\infty$ conditions. Additionally, the rarefaction parameter ($W_r^T$) and free-molecular bridging function ($Br$) capture the correct scaling of $C_f$ and $C_p$ at high $Kn_\infty$ and $M_\infty$. For the second comparison, the effect of partial surface accommodation is modeled with $\sigma_t = 0.4$ and $\sigma_n = 0.01$. It is important to note that the normal momentum accommodation coefficient ($\sigma_n$) is not equivalent to the normal energy accommodation coefficient ($\alpha_n$) which is used in the CL model; however, since the value of $\alpha_n$ is nearly specular in the DSMC simulations, $\sigma_n$ and $\alpha_n$ will be approximately equal. Comparisons of the $C_f$ and $C_p$ distributions with partial surface accommodation are shown in figures 11 and 12, respectively. In general, there is good agreement between the proposed models and the DSMC data. Due to the reduced tangential momentum exchange between gas particles and the surface of the sphere, the skin friction coefficient decreases relative to the fully diffuse condition. Additionally, because the particle's incoming momentum normal to the surface is reversed for specular reflections, the surface pressure increases relative to the fully diffuse case. In both scenarios, the proposed slip and rarefaction correction factors accurately capture the effect of partial surface accommodation across the simulated conditions.

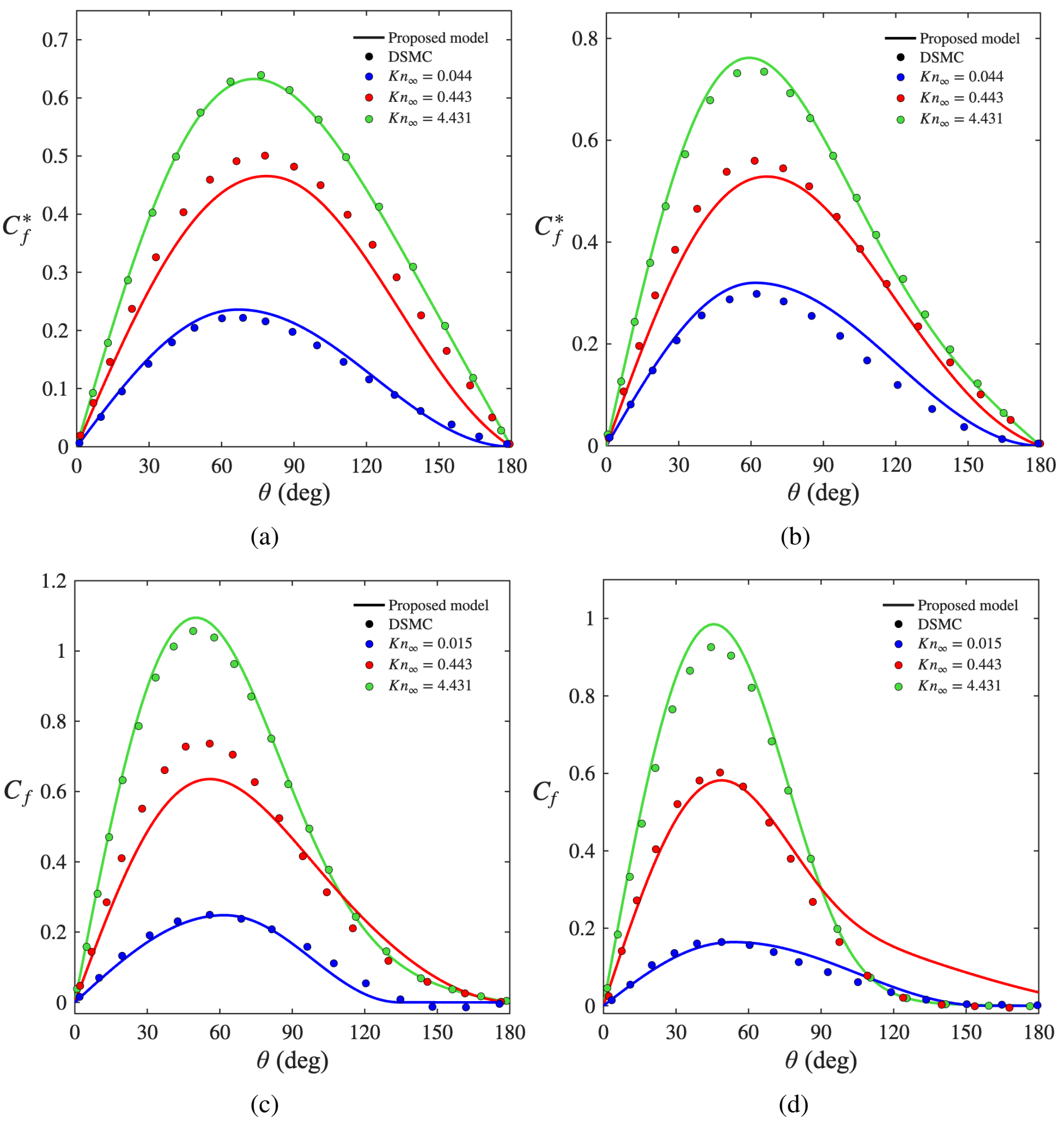

Figure 9: Comparison of $C_f$ distributions in rarefied flow with fully diffuse surface accommodation: (a) $M_\infty = 0.2$; (b) $M_\infty = 0.5$; (c) $M_\infty = 1$; (d) $M_\infty = 2$ ($C_f^* = C_f M_\infty$).

## 7. Conclusions

A new physics-based model for the distribution of shear stress and surface pressure is developed for laminar flow over spheres. The model depends on the flow regime of the sphere, including the freestream Reynolds ($Re_\infty$), Mach ($M_\infty$), and Knudsen ($Kn_\infty$) numbers. Additionally, it is a function of the local gas properties and surface conditions of the sphere, including the ratio of specific heat capacities ($\gamma$), viscosity power-law exponent ($\omega$), Prandtl number ($Pr$), surface temperature ratio ($T_w/T_0$), and surface accommodation coefficients ($\sigma_t$, $\sigma_n$). In the continuum regime, the model is formulated by considering boundary-layer scaling, flow separation, and shock-wave physics while recovering the known creeping-flow and hypersonic limiting behavior. To extend the model to the rarefied regime, the effects of velocity slip, gas-surface accommodation, and high-speed rarefaction are incorporated. Finally, to recover the analytical expressions for the local shear stress and surface pressure in the free-molecular limit, a drag coefficient based bridging function is used.

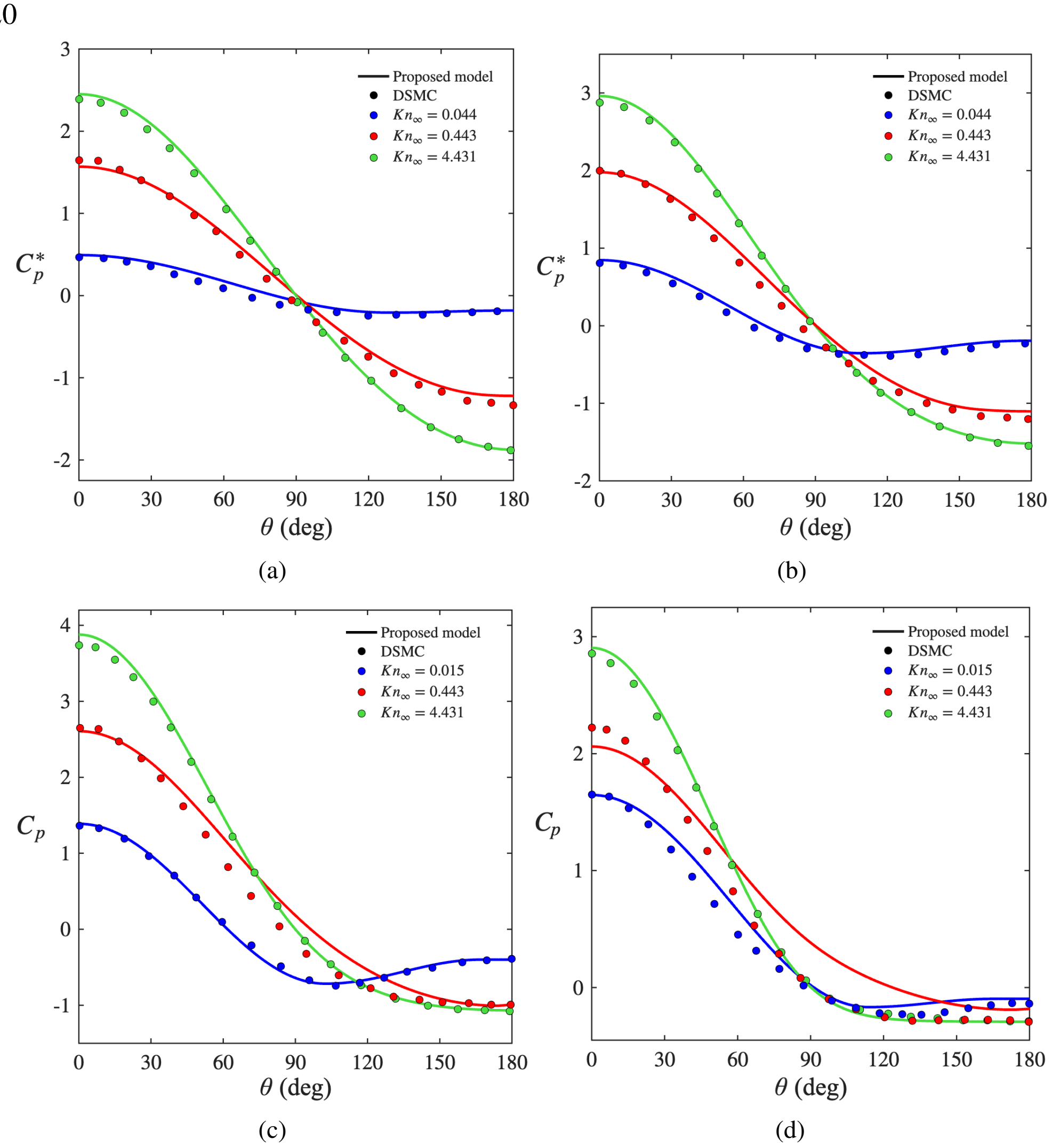

Figure 10: Comparison of $C_p$ distributions in rarefied flow with fully diffuse surface accommodation: (a) $M_\infty = 0.2$; (b) $M_\infty = 0.5$; (c) $M_\infty = 1$; (d) $M_\infty = 2$ ($C_p^* = C_p M_\infty$).

Model free parameters, all in the continuum regime, are fit to CFD data from literature. In the continuum regime, the angular dependence of the proposed model is validated using fully resolved CFD distributions. The model is compared with CFD simulations of flow over a sphere in air under subsonic, supersonic, and hypersonic conditions with varying surface temperatures. It captures the CFD distributions very well over a wide range of $Re_\infty$, $M_\infty$, and $T_w/T_0$. In the rarefied regime, the use of the rarefaction correction factors and bridging function is validated with fully resolved DSMC distributions in slip and transitional flow. The model is compared with DSMC simulations of flow over a sphere in helium under subsonic and supersonic conditions. It shows good agreement with the simulated distributions of shear stress and surface pressure over a wide range of $M_\infty$ and $Kn_\infty$ while accurately capturing the effects of fully diffuse and partial surface accommodation.

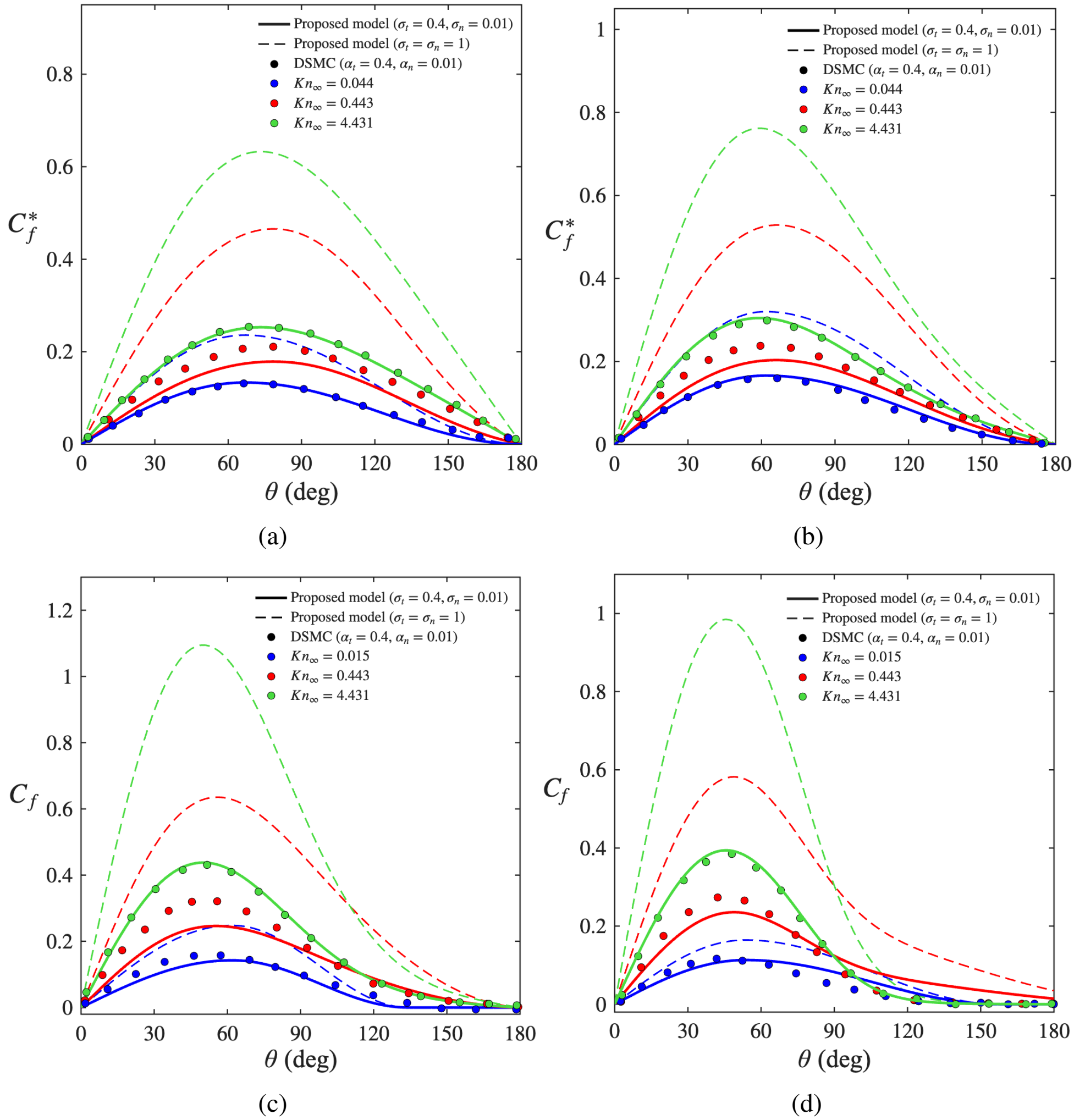

Figure 11: Comparison of $C_f$ distributions in rarefied flow with partial surface accommodation: (a) $M_\infty = 0.2$; (b) $M_\infty = 0.5$; (c) $M_\infty = 1$; (d) $M_\infty = 2$ ($C_f^* = C_f M_\infty$).

The new analytical models may be useful for simulating flows with particles or droplets where multiphase effects can depend on local properties around the particle surface. In particular, the distributions of the shear stress and surface pressure can provide the local surface conditions relevant to heat transfer, phase change, and mass loss.

**Acknowledgments.**

This research was partially supported by ONR Multidisciplinary University Research Initiative (MURI) grant #N00014-20-1-2682, and by the Office of the Under Secretary of Defense of Research and Engineering under award number FA9550-22-1-0342 with Dr.

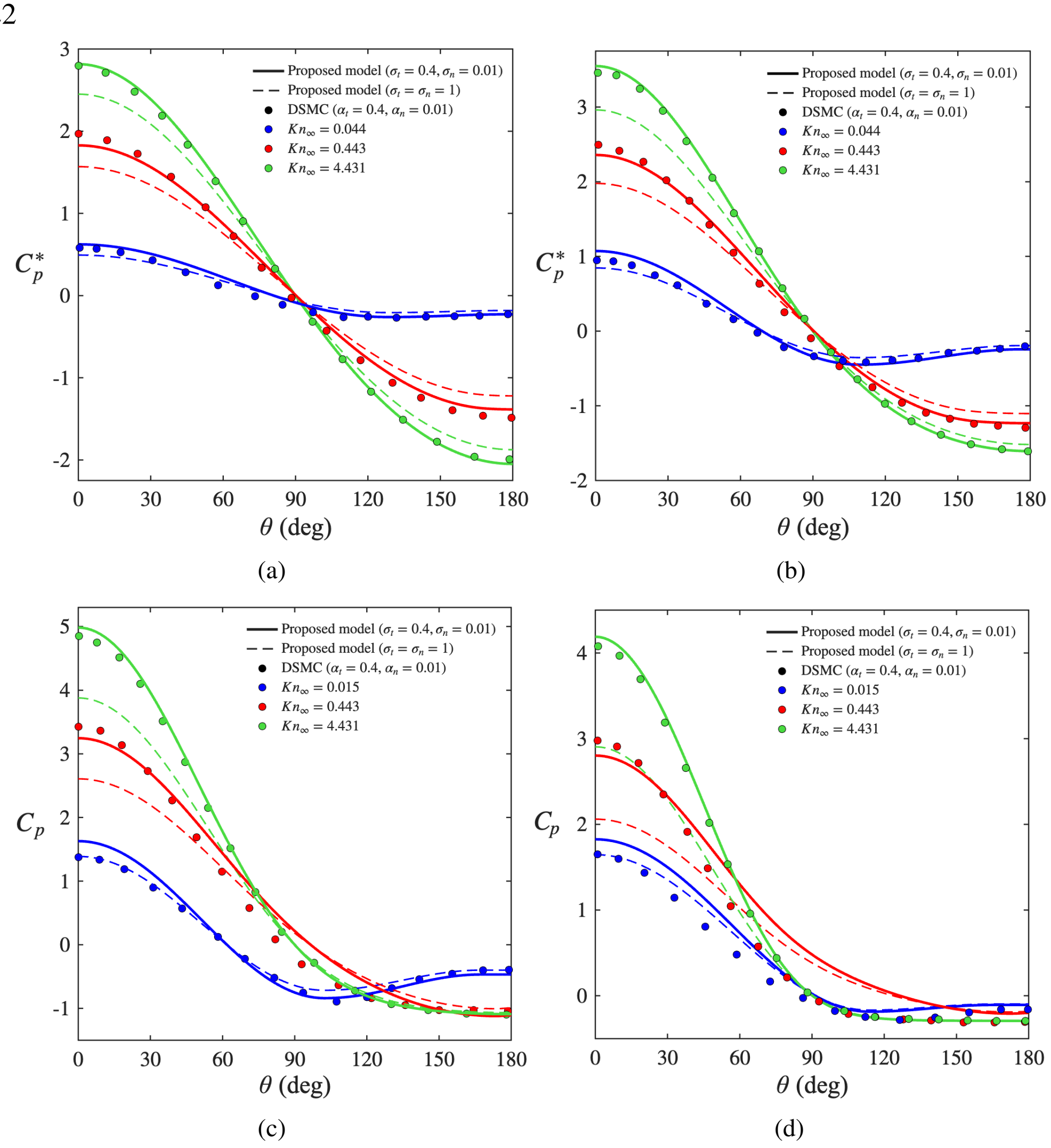

Figure 12: Comparison of $C_p$ distributions in rarefied flow with partial surface accommodation: (a) $M_\infty = 0.2$; (b) $M_\infty = 0.5$; (c) $M_\infty = 1$; (d) $M_\infty = 2$ ($C_p^* = C_p M_\infty$).

Eric Marineau as the program manager. This work was also supported by AFOSR grant FA9550-25-1-0302 with Dr. Amanda Chou as the program manager.

## Appendix A. Complete set of model equations

The full set of model equations for computing the distributions of $C_f$ and $C_p$ are summarized in this section. The expressions are separated based on the corresponding flow regime. Expressions that contain the variables $Re$, $M$, or $Kn$ without subscripts should be computed based on the freestream ($\infty$) or post-shock ($s$) values for subsonic and supersonic flow conditions, respectively. In general, $Kn$ can be related to $Re$ and $M$ with

$$Kn = \frac{M}{Re}\sqrt{\gamma\pi/2} \tag{A 1}$$

For $M_\infty > 1$, the post-shock variables can be computed with

$$Re_s = Re_\infty \left[1 + \frac{2(\gamma-1)}{(\gamma+1)^2 M_\infty^2}\left(M_\infty^2 - 1\right)\left(1+\gamma M_\infty^2\right)\right]^{-\omega} \tag{A 2}$$

$$M_s = \left(\frac{(\gamma-1)M_\infty^2 + 2}{2\gamma M_\infty^2 - (\gamma-1)}\right)^{1/2} \tag{A 3}$$

Continuum regime ($C_f$):

$$\begin{aligned}
C_f^c &= C_{f,\max} \sin\left(\frac{\pi}{2}\frac{\theta}{\theta_{\max}}\right) && 0^{\circ} \leqslant \theta \leqslant \theta_{\max} \\
C_f^c &= C_{f,\max} \sin^2\left(\frac{\pi}{2}\frac{\theta-\theta_s}{\theta_{\max}-\theta_s}\right) && \theta_{\max} < \theta \leqslant \theta_s \\
C_f^c &= 0 && \theta_s < \theta \leqslant 180^{\circ}
\end{aligned} \tag{A 4}$$

where

$$C_{f,\max} = \frac{6}{Re}\left(1+\frac{Re}{2.493}\right)^{0.5}\left(\frac{T^*}{T}\right)^{\omega/2} F_M(M) \tag{A 5}$$

$$\frac{T^*}{T} = 0.285 - 0.126\sqrt{Pr} + \left(0.715\frac{T_w}{T_0} + 0.126\sqrt{Pr}\right)\left(1+\frac{\gamma-1}{2}M^2\right) \tag{A 6}$$

$$F_M = 1 + 0.143 M_\infty^2 - 0.243 M_\infty^4 \quad M_\infty < 1 \tag{A 7$a$}$$

$$F_M = 0.56 + 0.351 M_s^2 - 0.011 M_s^4 \quad M_\infty \geqslant 1 \tag{A 7$b$}$$

Continuum regime ($C_p$):

$$\begin{aligned}
C_p^c &= C_{p,\min} + \left(C_{p,0} - C_{p,\min}\right)\cos^2\left(\frac{\pi}{2}\frac{\theta}{\theta_{\min}}\right) && 0^{\circ} \leqslant \theta \leqslant \theta_{\min} \\
C_p^c &= C_{p,\min} + \Delta C_{p,r}\cos^2\left(\frac{\pi}{2}\frac{\theta-\theta_{\mathrm{b}}}{\theta_{\min}-\theta_{\mathrm{b}}}\right) && \theta_{\min} < \theta \leqslant \theta_{\mathrm{b}} \\
C_p^c &= C_{p,\mathrm{b}} && \theta_{\mathrm{b}} < \theta \leqslant 180^{\circ}
\end{aligned} \tag{A 8}$$

where

$$C_{p,0} = C_{p,0,M} + \frac{12}{Re + 0.644\sqrt{Re}} \tag{A 9}$$

$$C_{p,0,M} = \frac{2}{\gamma M_\infty^2}\left[\left(1+\frac{\gamma-1}{2}M_\infty^2\right)^{\frac{\gamma}{\gamma-1}} - 1\right] \quad M_\infty < 1 \tag{A 10$a$}$$

$$C_{p,0,M} = \frac{2}{\gamma M_\infty^2}\left[\left(\frac{2\gamma M_\infty^2-(\gamma-1)}{\gamma+1}\right)\left(1+\frac{\gamma-1}{2}M_s^2\right)^{\frac{\gamma}{\gamma-1}}-1\right] \quad M_\infty \geqslant 1 \tag{A 10b}$$

$$C_{p,\min} = -\frac{6}{Re_\infty}\left(1+\frac{Re_\infty^{2.5}}{270}\right)^{0.384}\frac{1+0.174\min(1,M_\infty^2)+0.045\min(1,M_\infty^4)}{\max(1,M_\infty^2)} \tag{A 11}$$

$$\Delta C_{p,r} = \frac{(Re_\infty/20)^{1.586}}{1+168.5(Re_\infty/270)^{1.72}}\frac{1+0.23\min(1,M_\infty^2)-0.452\min(1,M_\infty^4)}{\max(1,M_\infty^2)} \tag{A 12}$$

$$C_{p,\mathrm{b}} = C_{p,\min}+\Delta C_{p,r} \tag{A 13}$$

Slip regime:

$$\beta_\tau = \frac{1}{1+6c_sKn(1+0.12Re)^{0.5}} \tag{A 14}$$

$$\beta_p = \frac{1+12c_sKn}{1+6c_sKn} \tag{A 15}$$

where

$$c_s = \frac{2-\sigma_t}{\sigma_t} \tag{A 16}$$

Transition regime:

$$\psi_{Wr,\tau} = \frac{\beta_\tau\phi_\tau}{1+1.27W_r^T} \tag{A 17}$$

$$\psi_{Wr,p} = \frac{\beta_p\phi_p}{1+1.27W_r^T} \tag{A 18}$$

where

$$\phi_\tau = \frac{1+c_s\sigma_tKn_\infty}{1+Kn_\infty} \tag{A 19}$$

$$\phi_p = \frac{1}{1+3Kn_\infty(1-0.1073\sigma_n)^{-1}} \tag{A 20}$$

$$W_r^T = \frac{M_\infty^{2\omega}}{Re_\infty}\left(1+\frac{T_w}{T_0}\right)^\omega \tag{A 21}$$

Free-molecular regime:

$$C_f^{fm} = \frac{\sigma_t\sin\theta}{s\sqrt{\pi}}\left[e^{-\xi^2}+\xi\sqrt{\pi}\,(1+\mathrm{erf}(\xi))\right] \tag{A 22}$$

$$C_p^{fm} = \frac{1}{s^2}\left[\left(\frac{2-\sigma_n}{\sqrt{\pi}}\xi + \frac{\sigma_n}{2}\sqrt{\frac{T_w}{T_\infty}}\right)e^{-\xi^2} + \left\{(2-\sigma_n)\left(\xi^2+\frac{1}{2}\right) + \frac{\sigma_n}{2}\xi\sqrt{\pi}\sqrt{\frac{T_w}{T_\infty}}\right\}(1+\mathrm{erf}(\xi)) - 1\right] \tag{A 23}$$

where

$$s = M_\infty\sqrt{\gamma/2} \tag{A 24}$$

$$\xi = s\cos\theta \tag{A 25}$$

Final expressions for $C_f$ and $C_p$:

$$C_f = C_f^c\frac{\psi_{W_r,\tau}}{1+Br^\eta} + C_f^{fm}\frac{Br^\eta}{1+Br^\eta} \tag{A 26}$$

$$C_p = C_p^c\frac{\psi_{W_r,p}}{1+Br^\eta} + C_p^{fm}\frac{Br^\eta}{1+Br^\eta} \tag{A 27}$$

where $\eta = 1.8$ and

$$Br = W_r^T\frac{M_\infty^{2\omega-1}+1}{M_\infty^{2\omega-1}} \tag{A 28}$$

Finally, the angular positions for computing $C_f$ and $C_p$ are summarized for incompressible, subsonic, and supersonic flow in the continuum regime. The expressions recover the correct asymptotic behavior in the Stokes and hypersonic flow limits, which are denoted by $\theta_{Stk}$ and $\theta_\infty$, respectively.

$$\theta = k_0 + \frac{\theta_{Stk} - k_0}{(1+k_1 Re_\infty)^{k_2}} \qquad M_\infty < 0.3 \tag{A 29a}$$

$$\theta = \theta^{ic}\left(1 + k_3 M_\infty^2 + k_4 M_\infty^4\right) \qquad 0.3 \leqslant M_\infty \leqslant 1 \tag{A 29b}$$

$$\theta = \theta_\infty\left[1 + \frac{k_5(M_\infty-1)}{1+k_6(M_\infty-1)^{k_7}}\right]\left[\frac{\theta_{M_\infty=1}}{\theta_\infty}\right]^{M_\infty^{-k_8}} \qquad M_\infty \geqslant 1 \tag{A 29c}$$

where $\theta_{M_\infty=1} = \theta^{ic}(1+k_3+k_4)$. The coefficients for computing $\theta$ in incompressible and compressible flow are provided in tables 1 and 2, respectively. To ensure the angular positions do not exceed the trailing edge of the sphere in compressible flow, the final value should be evaluated with $\theta = \min(180^{\circ}, \theta)$. In incompressible flow, the separation angle ($\theta_s^{ic}$) is computed based on the expressions of Clift, Grace & Weber (1978), which are given by

$$\begin{aligned} \theta_s^{ic} &= 180^{\circ} - 42.5^{\circ}\ln\left(\frac{Re_\infty}{20}\right)^{0.483} && 20 < Re_\infty \leqslant 400 \\ \theta_s^{ic} &= 77.84^{\circ} + 275^{\circ} Re_\infty^{-0.37} && 400 < Re_\infty \leqslant 3\text{x}10^5 \end{aligned} \tag{A 30}$$

| | $k_0$ | $k_1$ | $k_2$ | $\theta_{Stk}$ |
|---|---|---|---|---|
| $\theta^{ic}_{\text{max}}$ | 49.5$^\text{o}$ | 0.625 | 0.50 | 90$^\text{o}$ |
| $\theta^{ic}_{\text{min}}$ | 72$^\text{o}$ | 0.25 | 0.65 | 180$^\text{o}$ |
| $\theta^{ic}_{\text{b}}$ | 91$^\text{o}$ | 0.0033 | 0.85 | 180$^\text{o}$ |

Table 1: Coefficients for computing the incompressible angular positions of $C_f$ and $C_p$.

| | $k_3$ | $k_4$ | $k_5$ | $k_6$ | $k_7$ | $k_8$ | $\theta_\infty$ |
|---|---|---|---|---|---|---|---|
| $\theta_{\text{max}}$ | 0.012 | 0.1291 | 0.0394 | 4.480 | 2.700 | 1.320 | $\theta^* + 5^\text{o}$ |
| $\theta_s$ | 0.012 | 0.0625 | 0.558 | 1.200 | 1.011 | 0.072 | 180$^\text{o}$ |
| $\theta_{\text{min}}$ | -0.0186 | 0.2338 | 0.476 | 1.640 | 1.660 | 0.150 | 90$^\text{o}$ |
| $\theta_{\text{b}}$ | 0.0107 | 0.0403 | 0.019 | 1.280 | 3.100 | 0.700 | 180$^\text{o}$ |

Table 2: Coefficients for computing the compressible angular positions of $C_f$ and $C_p$.

In hypersonic flow, the sonic point is computed as $\theta^* = 34^\text{o} + 40^\text{o}\left(\frac{\gamma-1}{\gamma+1}\right)$.

**Declaration of interests.** The authors report no conflict of interest.